\documentclass[aps,pra,amsmath,amssymb,floatfix,twocolumn,amsmath,superscriptaddress,twocolumn,nofootinbib,tighten,letterpaper]{revtex4}

\usepackage{amsfonts, relsize, color}
\usepackage{graphics}
\usepackage{graphicx}
\usepackage{subfigure}
\usepackage{hyperref}
\usepackage{amssymb}
\usepackage{mathrsfs}
\usepackage{comment}
\usepackage{lscape}
\usepackage{multirow}
\usepackage{hhline}
\usepackage{multirow,tabularx}
\usepackage{array}
\newcolumntype{P}[1]{>{\centering\arraybackslash}m{#1}}

\usepackage[english]{babel}
\usepackage[utf8]{inputenc}
\usepackage{fancyhdr}
\usepackage{braket}

\newcommand{\Sx}{\sin(k_x)}
\newcommand{\Sy}{\sin(k_y)}

\newcommand{\Cx}{\cos(k_x)}
\newcommand{\Cy}{\cos(k_y)}
\newcommand{\Cz}{\cos(k_z)}

\begin{document}

\title{Generalized triple-component fermions: Lattice model, Fermi arcs and anomalous transport}

\author{Snehasish Nandy}
 \affiliation{Max-Planck-Institut f\"{u}r Physik komplexer Systeme, N\"{o}thnitzer Stra. 38, 01187 Dresden, Germany}
 \affiliation{Department of Physics, Indian Institute of Technology Kharagpur, WB 721302, India}

\author{Sourav Manna}
 \affiliation{Max-Planck-Institut f\"{u}r Physik komplexer Systeme, N\"{o}thnitzer Stra. 38, 01187 Dresden, Germany}

\author{Dumitru C\u{a}lug\u{a}ru}
 \affiliation{Max-Planck-Institut f\"{u}r Physik komplexer Systeme, N\"{o}thnitzer Stra. 38, 01187 Dresden, Germany}
 \affiliation{Cavendish Laboratory, University of Cambridge, J. J. Thomson Avenue, Cambridge, CB3 0HE, United Kingdom}
 \affiliation{Department of Physics, Princeton University, Princeton, New Jersey, 08544, USA}

\author{Bitan Roy}
 \affiliation{Max-Planck-Institut f\"{u}r Physik komplexer Systeme, N\"{o}thnitzer Stra. 38, 01187 Dresden, Germany}
 \affiliation{Department of Physics, Lehigh University, Bethlehem, Pennsylvania, 18015, USA}

\date{\today}

\begin{abstract}
We generalize the construction of time-reversal symmetry-breaking triple-component semimetals, transforming under the pseudospin-1 representation, to arbitrary (anti-)monopole charge $2 n$, with $n=1,2,3$ in the crystalline environment. The quasiparticle spectra of such systems are composed of two dispersing bands with pseudospin projections $m_s=\pm 1$ and energy dispersions $E_{\bf k}=\pm \sqrt{ \alpha^2_n k^{2n}_\perp +v^2_z k^2_z}$, where $k_\perp=\sqrt{k^2_x+k^2_y}$, and one completely flat band at zero energy with $m_s=0$. We construct simple tight-binding models for such spin-1 excitations on a cubic lattice and address the symmetries of the generalized triple-component Hamiltonian. In accordance to the bulk-boundary correspondence, triple-component semimetals support $2 n$ branches of topological Fermi arc surface states and also accommodate a \emph{large} anomalous Hall conductivity (in the $xy$ plane), given by $\sigma^{\rm 3D}_{xy} \propto 2 n \times$ the separation of the triple-component nodes (in units of $e^2/h$). Furthermore, we compute the longitudinal magnetoconductivity, planar Hall conductivity, and magneto thermal conductivity in these systems, which increase as $B^2$ for sufficiently weak magnetic fields ($B$) due to the nontrivial Berry curvature in the medium. A generalization of our construction to arbitrary integer spin systems is also highlighted.   
\end{abstract}
           
\maketitle


\section{Introduction}

Energy branches available for electrons to occupy in solid state compounds (also known as bands) can often \emph{touch} each other at few isolated and specific points in the Brillouin zone~\cite{herring, dornhaus, RyuTeo, Barnevig_2016, kane-prb, hasan-review, armitage-review}. In the close proximity to the band-touching points, the system can be described in terms of emergent pseudospin degrees of freedom, with the distinct eigenvalues of the pseudo-spin projection representing different bands. Some well known examples of such gapless systems are Dirac and Weyl semimetals~\cite{armitage-review}. Respectively in these two systems, Kramers degenerate and non-degenerate valence and conduction bands, transforming under half-integer pseudospin representations, touch each other. Such special points act as defects or singularities in the reciprocal space. For example, pseudospin-1/2 Weyl nodes in three dimensions assume the texture of a hedgehog or anti-hedgehog and stand as sources or sinks of Abelian Berry curvature, respectively. Nonetheless, it is also conceivable to realize band touching points around which the system can be described in terms of arbitrary pseudospin-$s$ representation, where $s$ can be any half-integer or integer. In the simplest incarnation of such higher pseudospin system, the energy spectra are described in terms of $(s+1/2)$ effective Fermi velocities, when $s$ is a half-integer; a phenomena known as \emph{multifringence}~\cite{luttinger, kennett-3, Dora, Lan-1, liangfu, manes, cano-bernevig, neupert-hasan, grushin_optical, Chen, Wieder-Kane,Venderbilt,chen-fiete,kane-rappe}. By contrast, for integer $s$, energy spectra display $s$ effective Fermi velocities and a completely flat band (described by the trivial eigenvalue of the pseudospin projections)~\cite{Barnevig_2016, Dai_2016, Ding_2017, Chen_2017, Hasan_2017, Zhang_2017, Soluyanov_2016, Hasan1_2017, Chang_arXiv, Chen_2018, fulga_2017, manes, cano-bernevig, neupert-hasan, grushin_optical, Chen}. The present work is devoted to unveil some quintessential topological features of time-reversal symmetry-breaking semimetals, transforming under the pseudospin-1 representation, also known as \emph{triple-component semimetals}, within the framework of both effective low-energy as well as representative tight-binding models on a cubic lattice. In quantum materials, emergent pseudospin degrees of freedom can arise from specific admixtures of orbital and spin projections~\cite{Barnevig_2016,cano-bernevig,neupert-hasan}, which should be distinguished from the real spin of Weyl fermions, appearing in the context of high-energy physics~\cite{weyl-original}.

Irrespective of these details, the entire family of pseudospin-$s$ Dirac or Weyl fermions can be described by the following effective low-energy Hamiltonian
\begin{equation}~\label{Eq:LowenergyHamil_Intro}
H_{\rm s} ({\bf k})= {\bf d}({\bf k}) \cdot {\bf S}, 
\end{equation}
where momenta ${\bf k}$ are measured from the band-touching points, and ${\bf S}$ are three spin-$s$ matrices~\footnote{In this work we focus only on three-dimensional systems. For the lattice realization of time-reversal symmetry breaking spin-1 system in two dimensions, see D. Green, L. Santos, and C. Chamon, Phys. Rev. B {\bf 82}, 075104 (2010), for example.}. In the simplest realization of a pseudospin-$s$ system ${\bf d}({\bf k})= v \; {\bf k}$, where $v$ bears the dimension of the Fermi velocity. The energy spectra are then given by $\pm E_s({\bf k})$, where $E_s({\bf k})=v_s |{\bf k}|$, with $v_s=\left( 1/2, 3/2, \cdots, s \right) v$ (for half-integer $s$) or $v_s=\left( 0, 1, \cdots, s \right) v$ (for integer $s$). We here concentrate on integer pseudospin-$s$ systems.~\footnote{ We here neglect the particle-hole asymmetry of the form $S_0 (a + b \; {\bf k}^2)$, where $S_0$ is a (2s+1) dimensional idenity matrix, which is always present in any real materials, since it does not affect the topology of the bands. However, for $b \neq 0$ the completely flat, topologically trivial band becomes dispersive.} Even though the following discussion can be generalized to any integer value of $s$, for the sake of concreteness we restrict ourselves to pseudospin-1 systems. The electronic excitations in such a setup are also known as triple-component fermions (due to three energy bands), and we here consider such peculiar band touching in time-reversal symmetry-breaking systems. In what follows, such band-touching points are referred to as \emph{triple-component points} or \emph{nodes}. In time-reversal symmetric systems, triple component fermions arise in materials with space group symmetry 199 and 214, with Pd$_3$Bi$_2$S$_2$ and Ag$_2$Se$_2$Au, in particular, accommodating such unconventional gapless fermionic excitations~\cite{Barnevig_2016}. However, time-reversal symmetry breaking triple-component fermions still lack material realizations. Nonetheless, spin-1/2 Weyl fermions have recently been found in magnetic materials, such as Mn$_3$Sn~\cite{mn3sn} and Ti$_2$MnAl~\cite{ti2mnal}. Hence, time-reversal symmetry breaking triple-component semimetals can be found (at least in principle) in magnetically ordered systems. In the present manuscript, we unearth rich topological properties of spin-1 systems by focusing on both continuum and lattice based ``toy" models.~We hope that the present discussion will motivate the search for spin-1 triple-component fermions in time-reversal symmetry breaking systems. But, finding possible candidate materials by performing a complete space group and magnetic point group analyses goes beyond the scope of the present discussion. Nonetheless, various topological responses of the generalized triple-component fermions can be tested numerically from the proposed lattice models in Sec.~\ref{Sec:LatticeModel}. We now present a brief synopsis of our main results.

\subsection{Summary of results}

We show that the triple component points act as sources and sinks of the Abelian Berry curvature with charges $ \pm 2 \; s\; n$ (with $n=1$, when ${\bf d}({\bf k})= v \; {\bf k}$), respectively. In this work we generalize the construction of triple component fermions for arbitrary monopole charges $|2 n|$ (assuming $s=1$), with $n=1,2,3$ in the crystalline environment (see Sec.~\ref{Sec:spin1_General}). Therefore, even though crystalline systems impose a stringent restriction on $n$, namely $n \leq 3$, one can still explore the territory of large monopole charge and its ramifications on topological responses by focusing on systems, where the bands transform under large spin $s$ representation. For an arbitrary integer value of $n$, the spectra always accommodate one topologically trivial flat band, while the energies of two dispersive bands scales as $E_{\bf k} \sim k_z$ and $E_{\bf k} \sim k^{n}_\perp$ (assuming that the triple-component points are separated along the $k_z$ direction), where $k_\perp =\sqrt{ k^2_x+k^2_y}$. The integer topological invariant of the system is given by ${\mathcal N}=2 \; n$. We also present simple lattice realizations of the generalized triple-component fermions on a cubic lattice and show that the generalized triple component nodes possess discrete fourfold rotational or $C_4$ symmetry (see Sec.~\ref{Sec:LatticeModel}).

The integer topological invariant of triple-component semimetals ($\mathcal N$) manifests through $2 n$ copies of topologically protected Fermi arc surface states connecting two triple-component points (see Figs.~\ref{Fig:FermiArc_2D} and ~\ref{Fig:FermiArc_3D}). This observation is in accordance with the \emph{bulk-boundary correspondence}, discussed in Sec.~\ref{Sec:SurfaceStates}. The time-reversal symmetry-breaking topological triple-component systems also support a \emph{large} anomalous Hall conductivity in a plane perpendicular to the separation of the two triple-component points (see Sec.~\ref{Sec:AHE}). The anomalous Hall conductivity acquires its largest value, given by (in units of $e^2/h$)
\begin{eqnarray}
\sigma^{\rm 3D}_{xy, {\rm max}}&=&\frac{2 n}{2\pi} \times  \: \big(\text{separation of triple-component} \nonumber \\
&& \text{points in the momentum space} \big)
\end{eqnarray}
when the chemical potential is pinned at the band touching points. Such a large anomalous Hall conductivity solely arises from the underlying Berry curvature of the medium~\footnote{The flat band at zero-energy is topologically trivial and possesses net zero Berry curvature. Hence, it does not affect any topological response of this system. Addition of particle-hole asymmetry in Eq.~(\ref{Eq:LowenergyHamil_Intro}) makes such a completely flat band dispersive, but it still remains topologically trivial.}. The Berry curvature can also leave its signature on other transport quantities, when, for example, the system is placed in a \emph{weak} magnetic field $B$ (hence no Landau quantization in the system). Specifically, we here compute the (a) longitudinal magnetoconductivity, (b) planar Hall conductivity and (c) longitudinal magneto-thermal conductivity, within the framework of the semiclassical Boltzmann theory and show that these quantities \emph{increase} as $B^2$, in the weak magnetic field regime, see Sec.~\ref{Sec:semiclassical} and Fig.~\ref{Fig:PHC_results}. The enhancement of various components of the magneto-conductivity tensor possibly captures the imprint of the \emph{chiral anomaly} in triple-component semimetals, discussed in the quantum limit in Ref.~\cite{lepori}. 
\\

\subsection{Outline}

The rest of the paper is organized as follows. In the next section, we introduce the effective low-energy models for generalized triple-component fermions and compute their topological invariant. Section~\ref{Sec:LatticeModel} is devoted to the construction of generalized triple-component fermions from simple tight-binding models on a cubic lattice. The topologically protected Fermi arc surface states and anomalous Hall conductivity are respectively discussed in Sec.~\ref{Sec:SurfaceStates} and Sec.~\ref{Sec:AHE}. Signatures of the Berry curvature on magneto transport are discussed in Sec.~\ref{Sec:semiclassical}. Concluding remarks and discussions on related issues are presented in Sec.~\ref{Sec:summary}. Additional technical details are relegated to the Appendices.           
\\


\section{Generalized triple component fermions: Low-energy model}~\label{Sec:spin1_General}

\begin{figure*}[t!]
\subfigure[]{
\includegraphics[height=5.00cm]{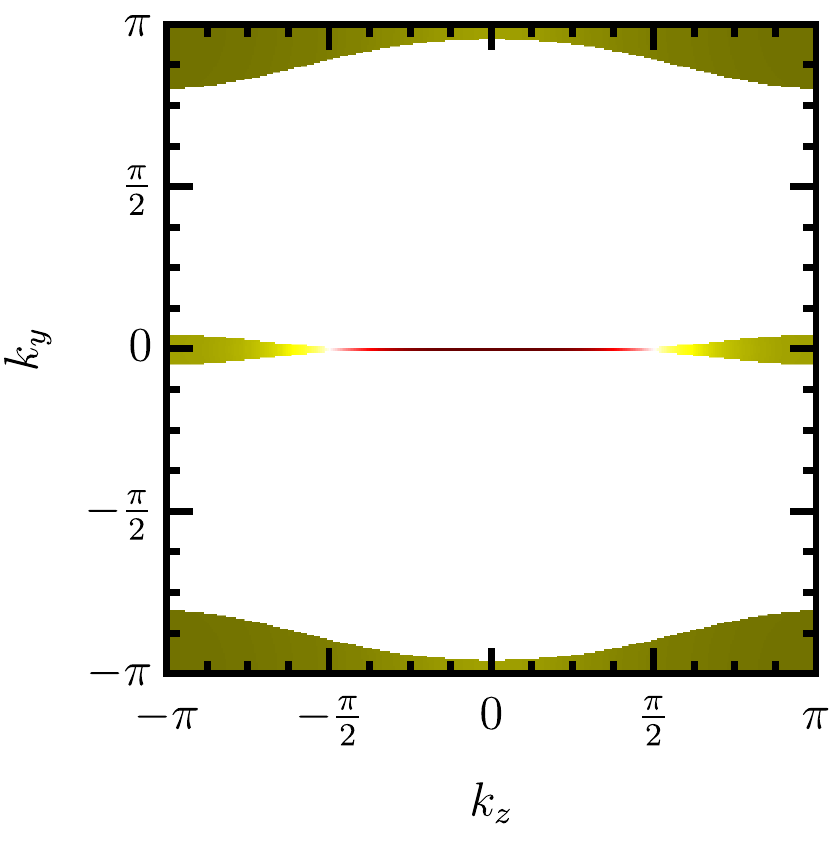}
}%
\subfigure[]{
\includegraphics[height=5.00cm]{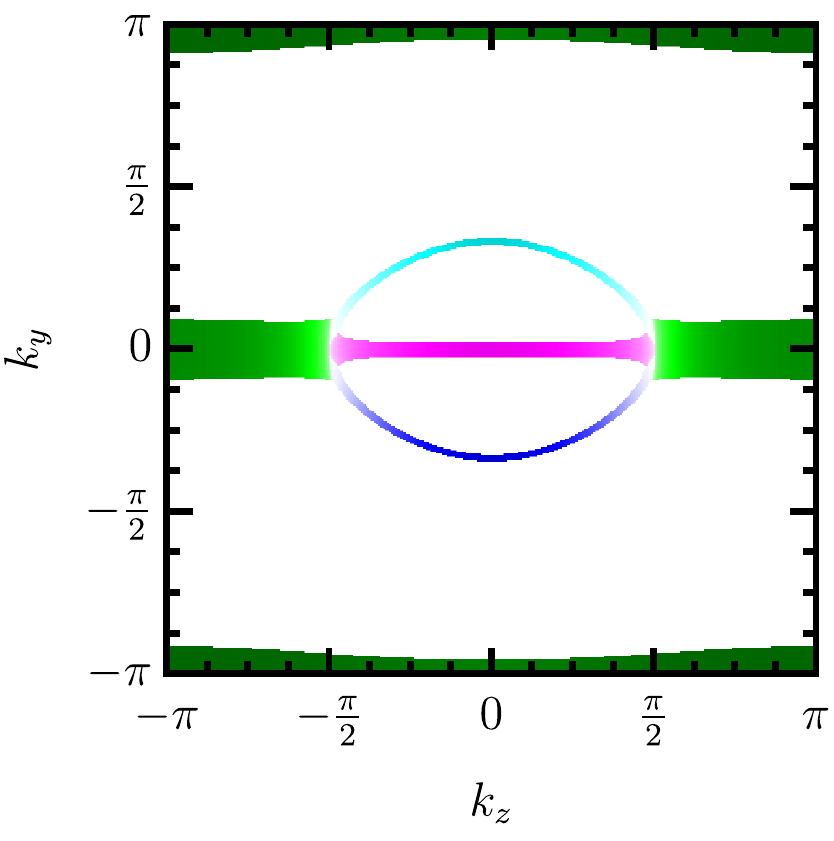}
}%
\subfigure[]{
\includegraphics[height=5.00cm]{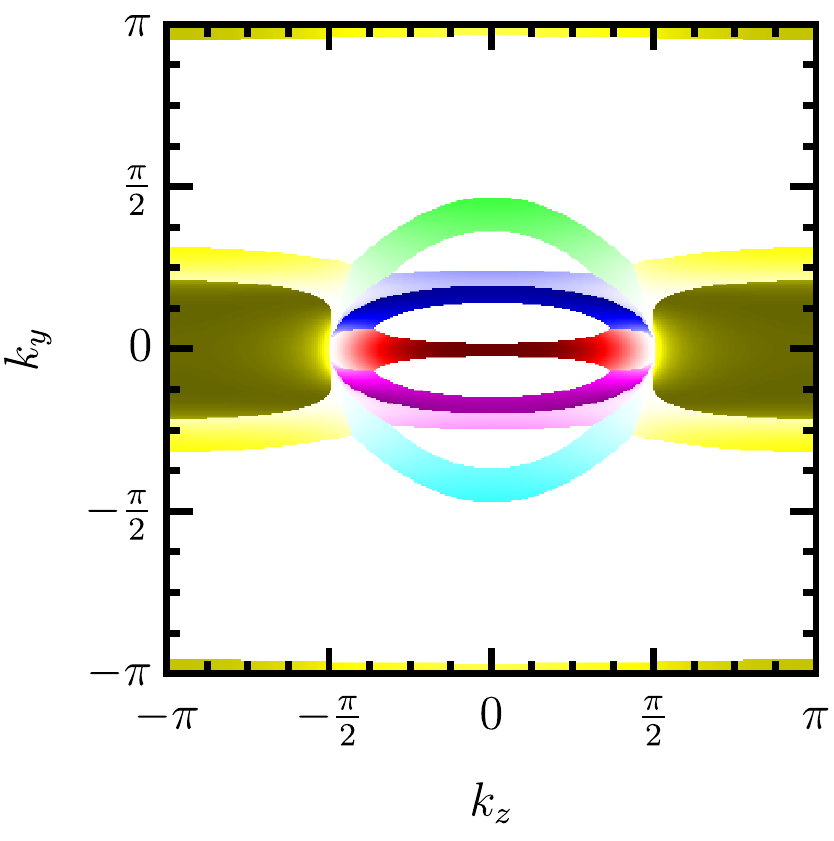}
}%
\includegraphics[height=5.00cm]{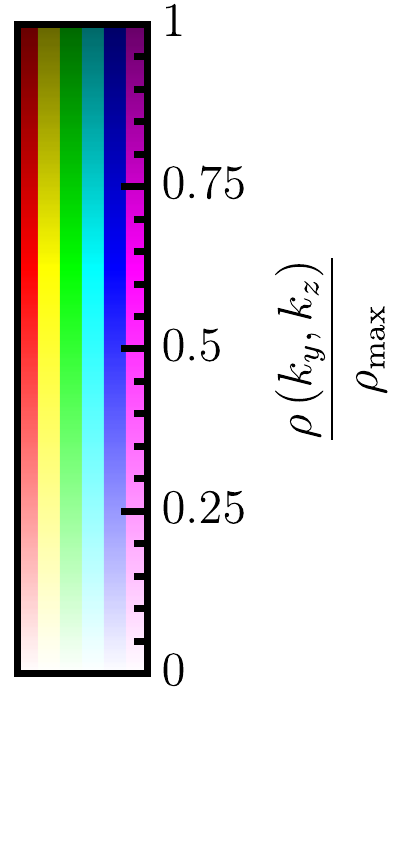}%
\caption{One-dimensional Fermi arc surface states in the $\left( k_y, k_z\right)$ plane for triple-component semimetals with (a) $n=1$, (b) $n=2$ and (c) $n=3$. We show the square of the amplitude $\rho(k_y,k_z)$  (normalized by its maximum value $\rho_{\rm max}$ to restrict within the interval $[0,1]$) of the low-energy wavefunctions that are localized on the (top) surface and reside within an energy window $\Delta E=0.02 t$ for (a), $0.04 t$ for (b) and $0.20 t$ for (c), around zero energy. Different branches of the Fermi arcs are shown in different colors. We impose periodic boundaries in the $y$ and $z$ directions (hence, $k_y$ and $k_z$ are good quantum numbers), but an open boundary in the $x$ direction, along which the linear dimension of the system is $L=280$. The triple component nodes are located at $\left( k_y, k_z \right)=(0, \pm \frac{\pi}{2})$ of the surface Brillouin zone. Note that a triple-component semimetal with monopole charge $2 n$ hosts $2 n$ branches of the Fermi arcs (see Sec.~\ref{Sec:SurfaceStates} for detailed discussion). Out of these, $(2n-1)$ branches directly connect the triple component points. One remaining branch is constituted by the surface localized state near $|k_y|=\pi$ and $0$. These seemingly disconnected pieces get connected once we increase the energy threshold $\Delta E$, see Appendix~\ref{Appendix:FermiArc_connected} and Fig.~\ref{Fig:FermiArc_connected}.    
}~\label{Fig:FermiArc_2D}
\end{figure*}

\begin{figure*}
\subfigure[]{
\includegraphics[height=6.00cm]{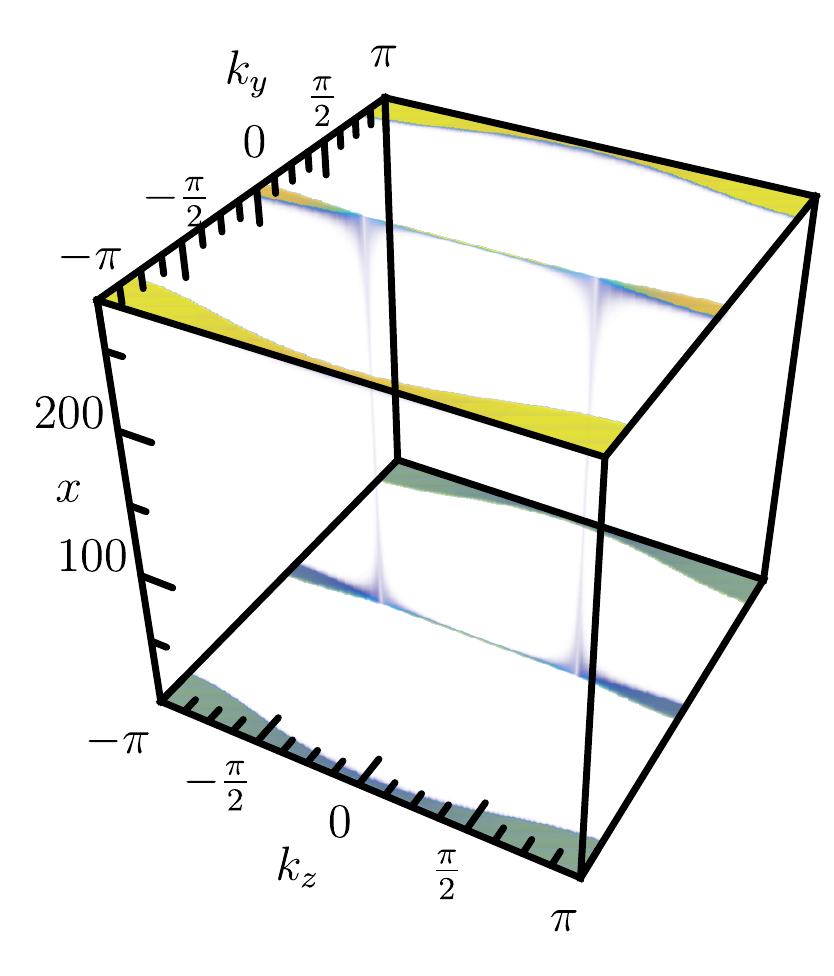}
}%
\subfigure[]{
\includegraphics[height=6.00cm]{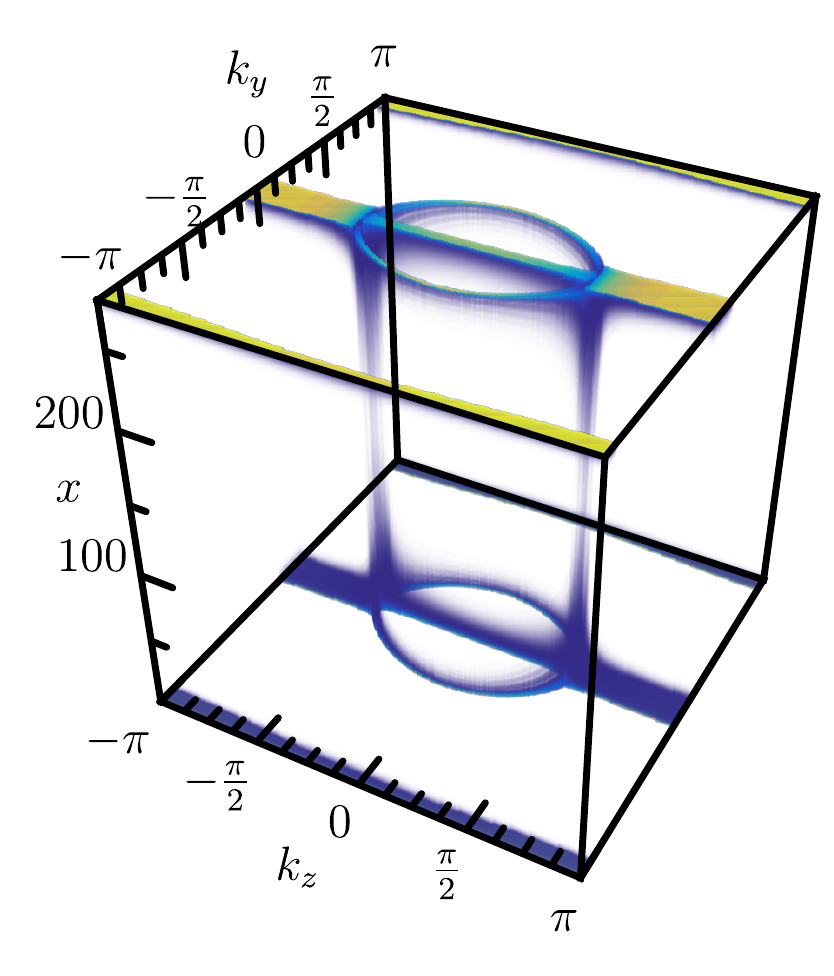}
}%
\subfigure[]{
\includegraphics[height=6.00cm]{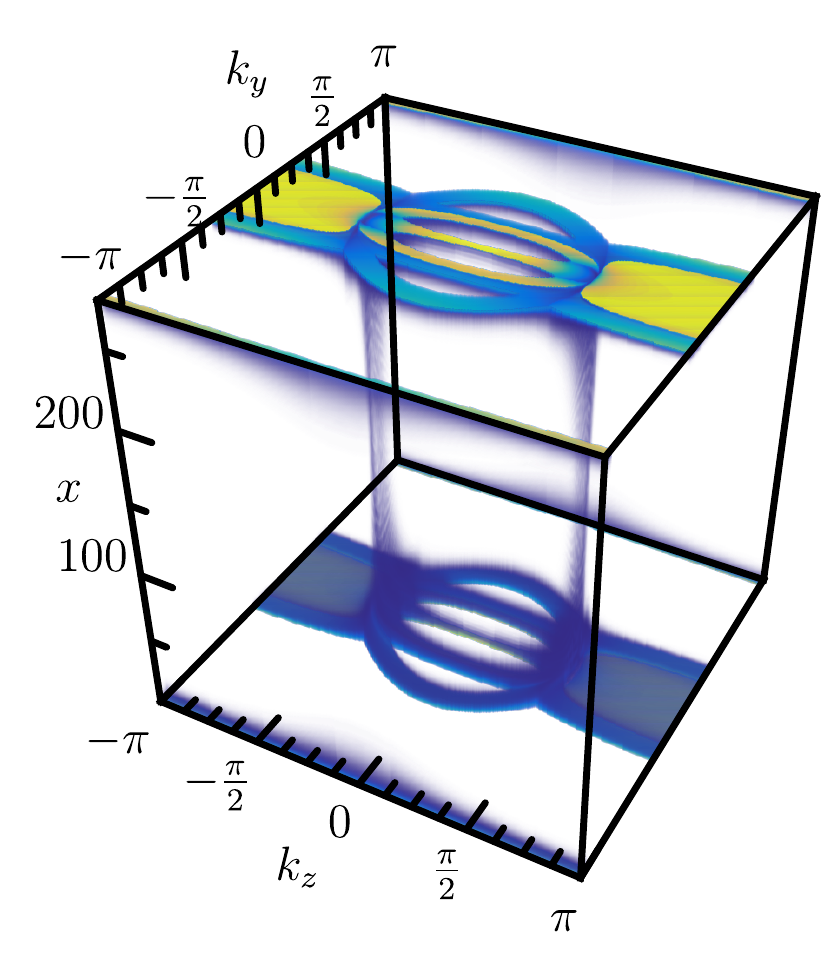}
}%
\includegraphics[height=6.00cm]{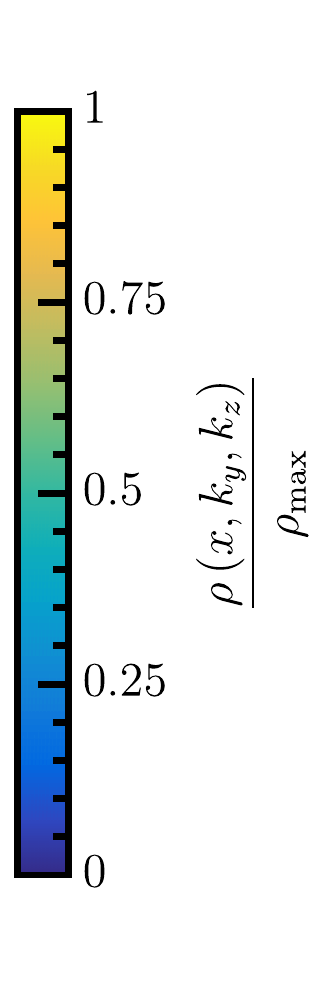}%
\caption{ Fermi arc surface states in three-dimensional mixed Bloch (along $k_y$ and $k_z$)-Wannier (along $x$) representation for triple-component semimetals with (a) $n=1$, (b) $n=2$ and (c) $n=3$. We display the same set of states shown in Fig.~\ref{Fig:FermiArc_2D} and implement the same boundary conditions. Notice that all branches of the Fermi arc surface states leak into the bulk through the triple-component nodes, located at $\left( k_y, k_z \right)=\left( 0, \pm \frac{\pi}{2}\right)$. Consequently, the top and bottom surfaces get connected through these two points (representing singularities in the momentum space), where the Fermi arcs are completely \emph{delocalized}. }~\label{Fig:FermiArc_3D}
\end{figure*}

We begin the discussion with the low-energy ``toy" models for general triple-component fermions (TCFs). The Hamiltonian operator describing such a system can be written compactly as 
\begin{equation}~\label{Hamiltonian_spin-1}
{\mathcal H}^{\rm TCF}_{n,\tau} ({\bf k})= d^n_1({\bf k}) \; S_x + d^n_2({\bf k}) \; S_y + \tau \; d_3({\bf k}) \; S_z, 
\end{equation}
where $S_x$, $S_y$ and $S_z$ are the spin-1 matrices, given by
\allowdisplaybreaks[4]
\begin{eqnarray}~\label{Eq:Spin1_matrices}
S_{x} &=& \frac{1}{\sqrt{2}}\begin{pmatrix} \begin{array}{ccc} 0 & 1 & 0 \\1 & 0 & 1\\0 & 1 & 0  \end{array} \end{pmatrix}
=\frac{1}{\sqrt{2}} \left( \lambda_1 + \lambda_6 \right), \nonumber \\
S_{y} &=& \frac{1}{\sqrt{2}}\begin{pmatrix} \begin{array}{ccc} 0 & -i & 0 \\i & 0 & -i\\0 & i & 0  \end{array} \end{pmatrix} 
=\frac{1}{\sqrt{2}} \left( \lambda_2 + \lambda_7 \right), \\
S_{z} &=& \begin{pmatrix} \begin{array}{ccc} 1 & 0 & 0 \\0 & 0 & 0\\0 & 0 & -1  \end{array} \end{pmatrix}
=\frac{1}{2} \left( \lambda_3 + \sqrt{3} \; \lambda_8 \right), \nonumber
\end{eqnarray} 
and $\lambda_j$s are the standard Gell-Mann matrices~\cite{Gell-Mann}. Here, $\tau=\pm$ represent the two valleys or triple-component points. Note that topological semimetals manifesting bulk-boundary correspondence (through Fermi arc surface states, see Figs.~\ref{Fig:FermiArc_2D} and \ref{Fig:FermiArc_3D}) must possess an even number of band touching points, giving rise to the notion of valley degrees of freedom, according to the Nielsen-Ninomiya theorem~\cite{nielsen-ninomiya}.
Different bands with energy dispersions $E^{m_s}_{\bf k}$ are characterized by distinct pseudospin projection $m_s=-1,0,1$. 
Energy spectra for TCFs in the vicinity of each valley are given by 
\begin{equation}
E^{m_s}_{\bf k}= m_s \sqrt{\left[ d^n_1({\bf k}) \right]^2+\left[ d^n_2({\bf k}) \right]^2 + d^2_3({\bf k})}.
\end{equation}
The bands with $m_s=\pm 1$, respectively describe the conduction and valence bands, and the flat band corresponds to $m_s=0$. Hence, TCFs accommodate two dispersive bands ($E^\pm_{\bf k}$) and one completely flat band ($E^0_{\bf k}$). We note that one can further generalize the above low-energy Hamiltonian from Eq.~(\ref{Hamiltonian_spin-1}) by taking
\begin{equation}
d_{3}({\bf k}) S_z \to d_{3}({\bf k}) \left[ S_z + \beta_{ij} N_{ij} \right],
\end{equation}
where the parameter $\beta_{ij}$ controls the coupling between the spin-tensor ($N_{ij}$) and momentum~\cite{spin-tensor}, with
\begin{equation} 
N_{ij}=\frac{1}{2} \left( S_i S_j + S_j S_i \right)-\frac{1}{3} \delta_{ij} {\bf S}^2.
\end{equation} 
Such spin-tensor to momentum coupling does not affect the topology of the system as long as $\beta_{ij} \ll 1$. Thus, for the sake of simplicity we set $\beta_{ij}=0$ throughout and work with the minimal model for spin-1 fermions introduced in Eq.~(\ref{Hamiltonian_spin-1}) and its lattice regularized version (see Sec.~\ref{Sec:LatticeModel}).

Next we discuss the nature of the dispersive bands for various choices of integer $n$. For any value of $n$, $d_3({\bf k})= v_z k_z$, whereas for $n=1$
\begin{equation}
d^1_1({\bf k})= \alpha_1 k_x, \:\: d^1_2({\bf k})= \alpha_1 k_y,
\end{equation}
where $\alpha_1$ bears the dimension of the Fermi velocity. We name the system \emph{linear triple-component semimetal}. For linear TCFs energy dispersions scale linearly with all the three components of momentum. By contrast, for $n=2$
\begin{equation}
d^2_1({\bf k})= \alpha_2 \left( k^2_x-k^2_y \right), \:\: d^2_2({\bf k})= \alpha_2 \left( 2 k_x k_y \right),
\end{equation}
and $\alpha_2$ bears the dimension of inverse mass. We name the system \emph{quadratic triple-component semimetal}. For quadratic TCFs, the dispersion scales linearly only with $k_z$, but quadratically with the in-plane components of momenta $k_x$ and $k_y$. Finally, for $n=3$ 
\begin{equation}
d^3_1({\bf k})= \alpha_3 \left( k^3_x-3 k_x k^2_y \right), \: d^2_2({\bf k})= \alpha_3 \left( k^3_y -3 k_yk^2_x \right),
\end{equation}
and we name this system \emph{cubic triple-component semimetal}. In this system the energy dispersion relation scales as $E^{\pm}_{\bf k} \sim k_{\perp}^3$. This construction can be envisioned as a generalization of the multi-Weyl systems composed of pseudospin-1/2 excitations~\cite{Fang-HgCrSe, bergevig-MWS, Hasan-PNAS, nagaosa, bera-roy-sau, ezawa-2} to the spin-1 systems. In Sec.~\ref{Sec:LatticeModel} we promote simple tight-binding models leading to the lattice realizations of such unconventional quasiparticles.

The topological charge of the triple-component points can be computed from the underlying Berry curvature of the bands. For simplicity, we now consider only one valley (say $\tau=+$). The Berry curvature of the $m$th band for a Bloch Hamiltonian [see Eq.~(\ref{Hamiltonian_spin-1})] is given by~\cite{Xiao:2010} 
\begin{eqnarray}
\Omega^{m}_{\theta \phi}=2i \sum_{m \neq m^{\prime}} \frac{\langle m | \frac{\partial H}{\partial \theta} |m^{\prime} \rangle \langle m^{\prime}|\frac{\partial H}{\partial \phi}|m \rangle}{(\epsilon_{m}-\epsilon_{m^{\prime}})^{2}},
\end{eqnarray}
where $\theta$ and $\phi$ are the polar and azimuthal angles in the momentum space, respectively and $m$, $m^\prime$ are the band indices. For convenience, we use the spherical coordinates. The wavefunctions for three bands read as 
\begin{eqnarray}
\bra{E^{-}_{\bf k}} &=& \left( e^{-2 i n \phi} \; \sin^2\frac{\theta}{2},\: -e^{-i n \phi} \; \frac{\sin\theta}{\sqrt{2}},\:\: \cos^2\frac{\theta}{2} \right), \nonumber \\
\bra{E^{+}_{\bf k}} &=& \left( e^{-2 i n \phi} \; \cos^2\frac{\theta}{2},\:\: e^{-i n \phi} \; \frac{\sin\theta}{\sqrt{2}},\:\: \sin^2\frac{\theta}{2} \right),  \\
\bra{E^{0}_{\bf k}} &=& \left( -e^{-2 i n \phi} \; \frac{\sin \theta}{\sqrt{2}},\:\: e^{-i n \phi} \; \cos\theta,\:\: \frac{\sin \theta}{\sqrt{2}} \right).\nonumber
\end{eqnarray}
The flat band possesses exactly \emph{zero} Berry curvature, while it is finite for the two dispersive bands. The monopole charge ($\mathcal N$) of the corresponding triple-component point can now be computed by integrating the Berry curvature over a unit sphere ($\rm A$) in the momentum space enclosing this point, yielding  
\begin{eqnarray}\label{Eq:MonopoleCharge}
{\mathcal N} = \frac{1}{2\pi}\int_{\rm A} d\mathbf{A} \cdot \mathbf{\Omega} =\frac{n}{2\pi} \int^\pi_0 \sin \theta \: d\theta  \int^{2 \pi}_0 d\phi
=2n.
\end{eqnarray}
The ``toy" models for spin-1 systems can be generalized for arbitrary integer ($s$) spin systems, for which the monopole charge is ${\mathcal N}=2 s n$. The monopole charge ${\mathcal N}$ also determines the integer topological invariant of such gapless phases of matter.


\section{Lattice models and symmetry}~\label{Sec:LatticeModel}

 We now propose simple tight-binding models that allow us to realize various members of the generalized TCF family on a cubic lattice. Such a simple lattice construction will also allow us to demonstrate the symmetries of triple-component points in the Brillouin zone, and topological features, such as (a) Fermi arc surface states (see Sec.~\ref{Sec:SurfaceStates}) and (b) the anomalous Hall effect (see Sec.~\ref{Sec:AHE}). For an arbitrary integer value of $n$, the corresponding tight-binding model takes a compact form 
\begin{equation}~\label{Eq:GeneralTCF_Lattice}
H = \sum_{\bf k} \Psi^\dagger_{\bf k} \:\: \left[ {\boldsymbol N}({\bf k}) \cdot {\boldsymbol S} \right] \:\: \Psi_{\bf k},
\end{equation}  
where $\Psi^\top_{\bf k} =\left[ c_{{\bf k},+1}, c_{{\bf k},0}, c_{{\bf k},-1}  \right]$ is a three-component spinor, $c_{{\bf k},s}$ is the fermion annihilation operator with momenta ${\bf k}$ and pseudo-spin projection $m_s=+1,0,-1$, and ${\bf S}=\left( S_x, S_y, S_z \right)$ [see Eq.~(\ref{Eq:Spin1_matrices})]. The momentum dependent form factors ${\bf N}({\bf k})$, appearing in Eq.~(\ref{Eq:GeneralTCF_Lattice}) for various values of $n$ arise from (setting the lattice constant $a=1$)~\cite{calugaru}
\begin{equation*}
 N_x({\bf k})= t \:
\begin{cases}
 \Sx & \text{for}\:\: n=1,\\
 \Cx -\Cy & \text{for}\:\: n=2,\\
 \Sx \left[ 3 \Cy-\Cx-2 \right] & \text{for}\:\: n=3,
\end{cases}
\end{equation*}
\begin{equation*} 
 N_y({\bf k})= t \:
\begin{cases}
 \Sy & \text{for}\:\: n=1,\\
 \Sx \Sy & \text{for}\:\: n=2,\\
 \Sy \left[ 3 \Cx-\Cy-2 \right] & \text{for}\:\: n=3,
\end{cases}
\end{equation*}
\begin{equation}~\label{Eq:LatticeModel_Components}
N^1_z({\bf k})=t_z \Cz,\:
N^2_z({\bf k})= m \left[ \Cx + \Cy -2 \right],
\end{equation}
with $N_z({\bf k})=N^1_z({\bf k})+N^2_z({\bf k})$. In this construction $N_x({\bf k})$ and $N_y({\bf k})$ give rise to the desired form factors for $d^n_x({\bf k})$ and $d^n_y({\bf k})$, respectively, when they are expanded around $\left( k_x, k_y \right)=(0,0)$ for $n=1,2,3$, when $t_z=m$. By contrast, $N^1_{z}$ produces two triple-component points at $k_z=\pm \frac{\pi}{2}$, whereas $N^2_z({\bf k})$ plays the role of a Wilson mass that only vanishes at $\left( k_x, k_y \right)=(0,0)$ for $t_z=m$. Therefore, with this construction, we end up with a minimal model for a time-reversal symmetry breaking general TCFs, for which the triple-component points are located at ${\bf k}=\left( 0,0,\pm \frac{\pi}{2}\right)$, see also Ref.~\cite{roy-slager-juricic}. The continuum ``toy" models discussed in Sec.~\ref{Sec:spin1_General} are realized from the above simple tight-binding models at low energies. We now highlight the symmetries of generalized triple-component nodes.

The generalized triple-component nodes possess discrete rotational symmetries. A rotation by an angle $\theta_{\rm PS}$ in the pseudo-spin space about its quantization axis (namely $S_z$) is captured by the unitary operator 
\begin{equation}
{\mathcal R}_{\rm PS} \left( \theta_{\rm PS}\right)=\exp \left[ i \; \theta_{\rm PS} \; S_z \right].
\end{equation}
On the other hand, a rotation of the momentum axis about the $k_z$ axis by an angle $\theta_{\bf k}$ is captured by 
\begin{equation}
{\mathcal R}_{\bf k} \left( \theta_{\bf k} \right)= \left( \begin{array}{ccc}
 \cos \theta_{\bf k} & -\sin \theta_{\bf k} & 0 \\
 \sin \theta_{\bf k} & \cos \theta_{\bf k} & 0 \\
  0 & 0 & 1
\end{array}
\right).
\end{equation}
Under a rotation by $\theta_{\rm PS}=n \; \frac{\pi}{2}$ in the pseudospin space 
\begin{eqnarray}
{\mathcal R}_{\rm PS} \left(\frac{n \pi}{2} \right) \; {\bf S}_\perp
{\mathcal R}^\dagger_{\rm PS} \left( \frac{n \pi}{2} \right) =
\begin{cases}
\left( -S_y, S_x \right)  & \text{for} \; n=1, \\
- \left( S_x, S_y \right) & \text{for}  \; n=2, \\
\left( S_y, -S_x \right) & \text{for}  \; n=3, 
\end{cases}
\end{eqnarray}
where ${\bf S}_\perp = (S_x, S_y)$. On the other hand, under the $C_4$ rotation by $\theta_{\bf k}=\frac{\pi}{2}$  
\begin{equation}
{\mathcal R}^\dagger_{\bf k} \left( \frac{\pi}{2} \right) \left( k_x, k_y, k_z \right) {\mathcal R}_{\bf k} \left( \frac{\pi}{2} \right)
= \left( -k_y, k_x, k_z \right).
\end{equation} 
Therefore, under a $C_4$ rotation in the momentum space, combined with a rotation by an angle $\theta_{\rm PS}=n \frac{\pi}{2}$ in the pseudospin space, the low-energy Hamiltonian ${\mathcal H}^{\rm TCF}_{n,\tau}$ from Eq.~(\ref{Hamiltonian_spin-1}) and its lattice regularized version, introduced in this section, remain invariant for $n=1$ and $2$. A similar argument in favor of the symmetry invariance of $n=2$ triple-component nodes has been recently reported in Ref.~\cite{zhu2018_doubeltriplecomponent}. Notice for $n=3$, both the terms proportional to $S_x$ and $S_y$ acquire an overall ``minus" sign. As a result, the topology of each node remains invariant under $C_4$ rotation. Alternatively, (a) one can define the rotation in the pseudospin space as $\theta_{\rm PS}=2 \pi-n \frac{\pi}{2}$, for $n=3$, which leaves the Hamiltonian for cubic triple-component fermions completely invariant, or (b) take $d^3_1 \leftrightarrow d^3_2$ in the continuum model or $N_x({\bf k}) \leftrightarrow N_y({\bf k})$ in the lattice model, such that they respectively read as 
\begin{eqnarray}
H^{\rm TCF}_{3,\tau}({\bf k}) &=& d^3_2({\bf k}) S_x + d^3_1({\bf k}) S_y + \tau d_3({\bf k}) S_z,  \\
H^{\rm TCF, lat}_{3,\tau}({\bf k}) &=& N_y ({\bf k}) S_x + N_x ({\bf k}) S_y + N_z ({\bf k}) S_z.
\end{eqnarray}
Both $H^{\rm TCF}_{3,\tau}({\bf k})$ and $H^{\rm TCF, lat}_{3,\tau}({\bf k})$ remain completely invariant under the above mentioned combined rotations in the momentum and pseudospin spaces. Therefore, the generalized triple-component nodes are invariant under $C_4$ rotations~\footnote{For any even $n$, the combined $C_4$ rotation in the momentum space and rotation by an angle $\theta_{\rm PS}=n \pi/2$ permit a unique power of $k_\perp$ in the low-energy Hamiltonian, namely $k^n_\perp$. Therefore, in a system with higher order band touching ($n>1$), a specific type of $k_\perp$-linear term, namely $({\bf S}_\perp \cdot {\bf k}_\perp)$, is forbidden by the $C_4$ symmetry~\cite{Fang-HgCrSe, bergevig-MWS, Hasan-PNAS}. Nonetheless, lattice distortion or external strain can reduce such symmetry and induce a $k_\perp$-linear term, which dominates when $k_\perp <k^\ast_\perp$, while the $k^n_\perp$ term dominates for $k_\perp >k^\ast_\perp$, where $k^\ast_\perp=(\alpha_1/\alpha_n)^{1/(n-1)}$. As long as $k^\ast_\perp \ll {\rm K}_0$, where ${\rm K}_0$ is the separation between the two Weyl nodes, the ultimate topological invariant of the system is given by $2 s n$. A somewhat similar situation occurs in two-dimensional monolayer and bilayer graphene, for example. Respectively in these two systems the coefficient of $k_\perp$ (namely, $\alpha_1$) and $k^2_\perp$ (namely, $\alpha_2$) dominates over the other one. In other words, $\alpha_1 \gg \alpha_2$ in monolayer, while $\alpha_1 \ll \alpha_2$ in bilayer graphene. Consequently, the topological invariants or the vorticities of these two gapless systems are $1$ and $2$, respectively~\cite{graphene-RMP}, despite the fact that the $k$-linear term splits an $n=2$ vortex in bilayer graphene into three vortices with $n=1$ and one antivortex with $n=-1$, such that the net vorticity around a given ${\rm K}$ point in the Brillouin zone remains as 2. By contrast, for odd $n$ the topology of the higher-order band touching ($\sim k^n_\perp$) remains unchanged as long as $\alpha_n \gg \alpha_1$, even though the $C_4$ symmetry permits $({\bf S}_\perp \cdot {\bf k}_\perp)$ term.}. Note that the above discussion on the symmetry does not rely on any specific value of $s$, and it is equally applicable to arbitrary integer and half-integer values of $s$. Next we demonstrate the bulk-boundary correspondence and construct the topological Fermi arc surface states by numerically diagonalizing the above tight-binding models in a cubic lattice for different $n$.


\section{Fermi arc surface states}~\label{Sec:SurfaceStates}

The hallmark of a topologically nontrivial phase of matter is the existence of surface or edge states, encoding the \emph{bulk-boundary correspondence}. The structure of such boundary modes, however, crucially depends on the actual nature of the bulk topological phase. For example, a three-dimensional strong topological insulator supports two-dimensional massless helical Dirac fermions on all six surfaces of a cubic system~\cite{hasan-kane-review, qi-zhang-review}. The surface modes of a topological semimetal are somewhat different from the ones of a topological insulator. Note that a time-reversal symmetry-breaking triple-component semimetal can be constructed by stacking two-dimensional layers of quantum anomalous Hall insulators of spin-1 fermions in the momentum space along the $k_z$ direction between two triple-component points, located at ${\bf k}=\left(0,0,\pm \frac{\pi}{2} \right)$~\footnote{Since the flat band is topologically trivial, we characterize each two-dimensional slice of the system as an ``insulator" even though there exists a dispersionless flat band at the middle of the band gap between the dispersive valence and conduction bands.}. Each copy of two-dimensional anomalous Hall insulator supports one-dimensional chiral edge modes, accommodating \emph{one} state precisely at zero energy. The collection of such zero-energy states between the two triple-component points constitute the \emph{topological Fermi arcs}, shown in Figs.~\ref{Fig:FermiArc_2D} and ~\ref{Fig:FermiArc_3D} for $n=1, 2, 3$. Such a seemingly hypothetical construction of time-reversal symmetry-breaking topological semimetals, nonetheless, leaves its signature in the anomalous Hall response of these systems, discussed in Sec.~\ref{Sec:AHE}.

In order to construct the Fermi arc surface states, we impose periodic boundaries in the $y$ and $z$ directions (hence leaving momenta $k_y$ and $k_z$ as good quantum numbers) and implement open boundary in the $x$ direction, along which the linear dimension of the system is set to be $L=280$.  For numerical diagonalization of the lattice models, we set $t=t_z=m=1$. In this construction, we can observe localized Fermi arcs states (in the mixed Wannier-Bloch representation) on the top and bottom surfaces~\cite{slager}. The resulting Fermi arcs for $n=1,2$, and $3$ are shown in Figs.~\ref{Fig:FermiArc_2D}(a),~\ref{Fig:FermiArc_2D}(b) and~\ref{Fig:FermiArc_2D}(c), respectively, and 
\begin{eqnarray}
\text{the number of Fermi arcs}= 2 \; n = \nonumber \\
\text{ monopole charge of triple-component points}.     
\end{eqnarray}
However, the counting of the Fermi arc states is a subtle issue. So it is worth pausing to illustrate it explicitly. Note that $(2n -1)$ copies of the Fermi arcs connect two triple component points located at $k_z=\pm \frac{\pi}{2}$ and are localized near the center of the surface Brillouin zone. The remaining copy of the Fermi arc surface states is constituted by the seemingly disconnected pieces, localized near $k_y=0,\pm \pi$, but spans the entire surface Brillouin zone along $k_z$, i.e. $-\pi \leq k_z \leq \pi$, see Figs.~\ref{Fig:FermiArc_2D} and ~\ref{Fig:FermiArc_3D}. These two pieces appear to be disconnected in Figs.~\ref{Fig:FermiArc_2D} and ~\ref{Fig:FermiArc_3D}, since we display the spectral weight of the surface localized states within specific energy windows. However, with increasing energy window these two pieces get gradually connected, see Appendix~\ref{Appendix:FermiArc_connected} and Fig.~\ref{Fig:FermiArc_connected}. The fact that this segment of the Fermi arc state extends beyond (despite being connected to) the bulk triple component points, is possibly specific for the simple lattice models we introduced in Sec.~\ref{Sec:LatticeModel} and the choice of the surface cut, namely the (100) plane. Nonetheless, Fermi arcs extending beyond the triple component points has also been noticed in Ref.~\cite{spin-tensor}, but only for $n=1$ and on the $(110)$ plane. However, only the segments localized within $-\frac{\pi}{2} \leq k_z \leq \frac{\pi}{2}$ contribute to the anomalous Hall conductivity, and it is precisely \emph{zero} for an underlying two-dimensional insulator in the $(k_x,k_y)$ plane when $k_z>\frac{\pi}{2}$ and $k_z<-\frac{\pi}{2}$ (see Sec.~\ref{Sec:AHE}). Hence the bulk-boundary correspondence remains operative for spin-1 semimetals, as there exist two, four and six branches of the Fermi arc surface states, respectively for $n=1, 2$, and $3$. Next we discuss some additional salient features of the arc states.

Note that the surface localization of the aforementioned $2n$ copies of the Fermi arc states is \emph{maximal} at its \emph{center} ($k_z=0$). As we approach the two triple-component nodes the surface localization decreases monotonically. At the two triple-component points the arc states become completely \emph{delocalized}, and the top and bottom surfaces get connected through the bulk triple-component points, as shown in Fig.~\ref{Fig:FermiArc_3D}. This observation does not depend on the choice of the integer value of $n$ and can be appreciated in the following way. Note that the localization length of the Fermi arc state for each value of $k_z$ is \emph{inversely} proportional to the size of the spectral gap of the corresponding two-dimensional layer of the quantum anomalous Hall insulator. From Eq.~(\ref{Eq:LatticeModel_Components}), one can appreciate that the bulk gap for the underlying two dimensional Hall insulating phase is \emph{largest} when $k_z=0$ (center of the arc). Otherwise, such a gap decreases monotonically as we approach the singular points, located at $k_z=\pm \frac{\pi}{2}$, from the center of the Fermi arcs, and \emph{vanishes} at $k_z=\pm \frac{\pi}{2}$. Consequently, the arc state at these two points become completely delocalized and two opposite surfaces get connected through the bulk triple-component points, which can be seen from Fig.~\ref{Fig:FermiArc_3D}. By contrast, the segments of the Fermi arc, localized near $|k_y|=\pi$, do not show any variation of surface localization along its length, since it is not connected to the triple-component points. Nonetheless, the part of the same Fermi arc, localized near $k_y=0$ displays variation of surface localization and leaks through the bulk triple component points, similarly to the remaining $(2n-1)$ copies of the Fermi arc. Hence, our numerical analyses of the Fermi arc surface states provide strong evidence in favor of the bulk-boundary correspondence in triple-component semimetals, with an arbitrary monopole charge $2 n$.

\begin{figure*}[t!]
\subfigure[]{ 
\includegraphics[height=4.0cm]{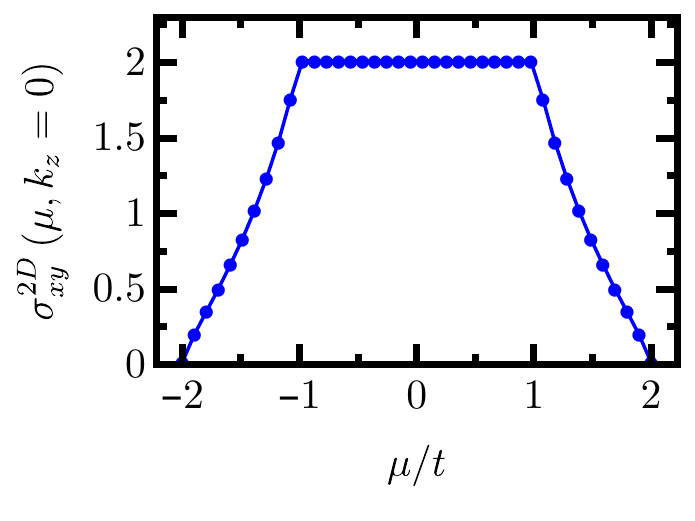}~\label{Fig:2DAHE_Simple}
}%
\subfigure[]{ 
\includegraphics[height=4.0cm]{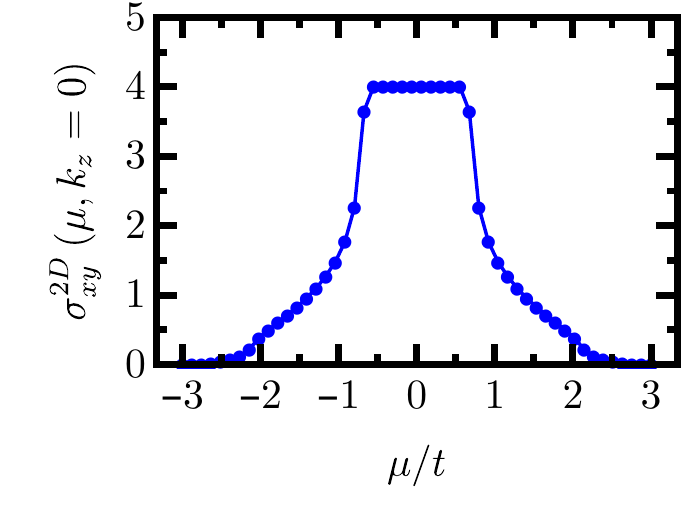}~\label{Fig:2DAHE_Double} 
}%
\subfigure[]{ 
\includegraphics[height=4.0cm]{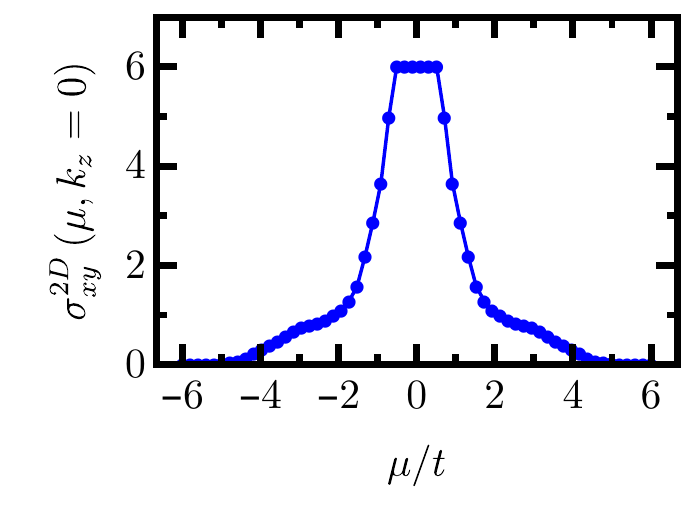}~\label{Fig:2DAHE_Triple} 
}
\subfigure[]{ 
\includegraphics[height=4.0cm]{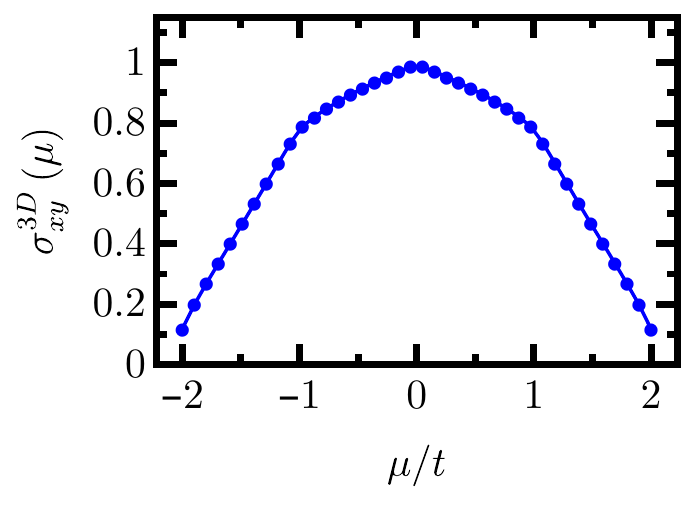}~\label{Fig:3DAHE_Simple} 
}%
\subfigure[]{ 
\includegraphics[height=4.0cm]{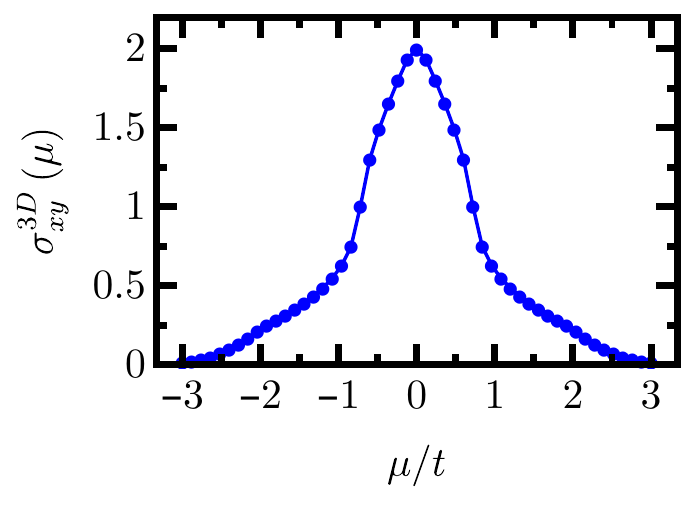}~\label{Fig:3DAHE_Double} 
}%
\subfigure[]{ 
\includegraphics[height=4.0cm]{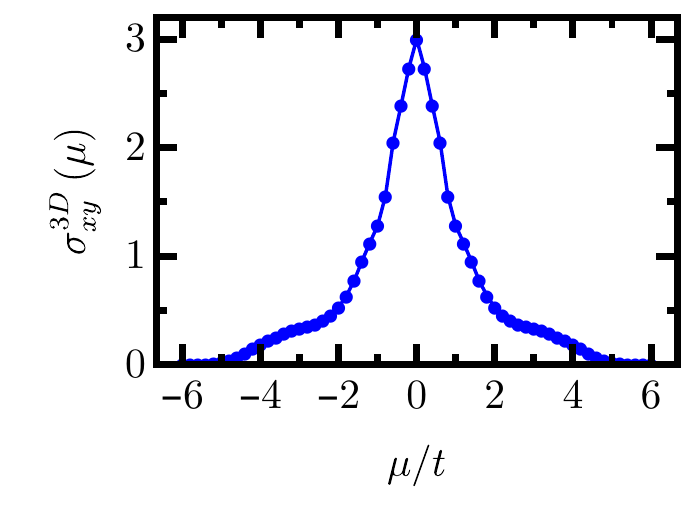}~\label{Fig:3DAHE_Triple} 
}%
\caption{Upper row: Scaling of the anomalous Hall conductivity (AHC), measured in units of $e^2/h$, given by Eq.~(\ref{Eq:AHE_2D_formula}), of an underlying constituting two dimensional layer of the time-reversal symmetry-breaking anomalous Hall insulator, obtained by setting $k_z=0$ in the tight-binding models for the triple-component semimetals, introduced in Eq.~(\ref{Eq:GeneralTCF_Lattice}) for (a) $n=1$, (b) $n=2$ and (c) $n=3$. The size of the insulating gap for anomalous Hall insulator of spin-1 fermions is $2t$. For chemical doping $|\mu|<t$, the AHC remains quantized. For $|\mu|>t$ the AHC decreases monotonically, and becomes zero when the system is either fully filled or completely empty. We arrive at qualitatively similar results for $-\frac{\pi}{2} <k_z < \frac{\pi}{2}$. But, the AHC identically vanishes (for any $\mu$) when $k_z>\frac{\pi}{2}$ and $k_z<-\frac{\pi}{2}$. Lower row: Corresponding AHC for three-dimensional triple-component topological semimetal with (d) $n=1$, (e) $n=2$ and (f) $n=3$, see Eq.~(\ref{Eq:AHE_3D_formula}) [in units of $e^2/(h a)$, where $a$ is lattice spacing, set to be unity for convenience]. Note that the AHC for the triple-component semimetals is largest when $\mu=0$, see Eq.~(\ref{Eq:AHE_3D_results}). With increasing or decreasing chemical doping away from the triple component band-touching points, the AHC decreases monotonically, and it vanishes when $\mu$ meets the band edges.        
}~\label{Fig:AHE_Together}
\end{figure*}

\section{Anomalous Hall Effect}~\label{Sec:AHE}

Yet another hallmark of a time-reversal symmetry-breaking topological semimetals is the nontrivial anomalous Hall effect in a plane perpendicular to the separation of two band touching points. We discussed in the previous section that both Weyl and triple-component semimetals can be constructed by stacking two-dimensional quantum anomalous Hall insulators in the momentum space along a specific (in our construction $k_z$) direction. Since each layer produces quantized (thus \emph{large}) anomalous Hall conductivity (AHC), the resulting time-reversal symmetry-breaking semimetallic phase also supports finite (typically large, but nonquantized) AHC.

For any value of $n$ the system describes an anomalous Hall insulator in the $xy$ plane that supports quantized AHC, given by 
\begin{equation}~\label{Eq:AHE_2D_results}
\sigma^{\rm 2D}_{xy} \left( \mu=0, k_z=0 \right)=\frac{e^2}{h} \times (2n).
\end{equation}  
The AHC can be directly computed from the underlying Berry curvature ($\Omega_z$)
\begin{equation}~\label{Eq:AHE_2D_formula}
\sigma^{\rm 2D}_{xy}(\mu, k_z=0)=\frac{e^2}{h} \int \frac{d^2{\bf k}_\perp}{(2 \pi)^2} \: \Omega_z f^0_{m_s}(\mu),
\end{equation} 
where $f^0_{m_s}(\mu)$ is the equilibrium Fermi-Dirac distribution at finite chemical doping $\mu$ (measured from zero energy, see Fig.~\ref{Fig:AHE_Together}), given by 
\begin{equation}
f^0_{m_s}(\mu)= \left[ 1+ \exp \left(\frac{E^{m_s}_{\bf k}-\mu}{k_B T} \right) \right]^{-1},
\end{equation}
where $m_s=1,0,-1$. Here the Boltzmann constant $k_B$ is set to be unity. A direct correspondence between the AHC and the first Chern number of the underlying anomalous Hall insulator, obtained from the continuum models of these systems is presented in Appendix~\ref{App:UVRegularization}. First, we compute the AHC for two-dimensional time-reversal symmetry breaking insulator for $n=1,2,3$ from the lattice model, shown in Sec.~\ref{Sec:LatticeModel}, upon setting $k_z=0$. The results are displayed in Fig.~\ref{Fig:AHE_Together}(a) for $n=1$, Fig.~\ref{Fig:AHE_Together}(b) for $n=2$ and Fig.~\ref{Fig:AHE_Together}(c) for $n=3$, as a function of varying chemical doping ($\mu$). We tune $\mu$ over the entire energy band, i.e. from the bottom of the valence band to the top of the conduction band.

For completely empty bands the Berry curvature from the conduction and valence bands cancel each other, and the system supports precisely \emph{zero} AHC. As one increases $\mu$ from the bottom of the valence band, AHC starts to increase monotonically and reaches its quantized value when $\mu$ meets the top of the valence band. Note that when the chemical potential is pinned within the bulk insulating gap the $\sigma^{\rm 2D}_{xy}$ remains constant and \emph{quantized}, given by Eq.~(\ref{Eq:AHE_2D_results}). On further increasing $\mu$, the AHC starts to decrease, as the Berry curvature from the fully filled valence band gets partially canceled by that from the partially filled conduction band. Ultimately, when the conduction band becomes fully occupied, the Berry curvatures from these two bands completely cancel each other and the AHC once again drops back to zero. Otherwise, this feature is common for $n=1,2,3$. Note that the topologically trivial flat band residing at zero energy does not influence the AHC in two dimensions. We arrive at the same results (qualitatively) for any $k_z$ residing within the range $-\frac{\pi}{2} \leq k_z \leq \frac{\pi}{2}$. However, for $k_z>\frac{\pi}{2}$ and $k_z <-\frac{\pi}{2}$, the AHC is always identically zero for any value of chemical doping $\mu$. This observation ensures that the surface localized states for $k_z>\frac{\pi}{2}$ and $k_z <-\frac{\pi}{2}$ shown in Figs.~\ref{Fig:FermiArc_2D} and \ref{Fig:FermiArc_3D} do not contribute to AHC, and Fermi arcs extending beyond the triple component points is purely an artifact of the simple lattice models and (possibly) the choice of the surface.

The AHC for a three-dimensional time-reversal symmetry-breaking triple-component semimetal ($\sigma^{\rm 3D}_{xy}$) can be obtained by accumulating the contributions from each constituting two-dimensional layer of anomalous Hall insulator and is given by~\cite{Balents_2017} 
\begin{equation}~\label{Eq:AHE_3D_formula}
\sigma^{\rm 3D}_{xy} (\mu)= \int^{\frac{{\rm K}_0}{2}}_{-\frac{{\rm K}_0}{2}} \:\:\: \frac{dk_z}{2 \pi} \:\: \sigma^{\rm 2D}_{xy}(\mu, k_z), 
\end{equation}   
where ${\rm K}_0$ is the separation of two triple-component points. Since for $\mu=0$ each copy of underlying two-dimensional anomalous Hall insulator yields the largest and quantized AHC, the three-dimensional triple-component semimetal also yields largest AHC, given by 
\begin{equation}~\label{Eq:AHE_3D_results}
\sigma^{\rm 3D}_{xy, {\rm max}}= \sigma^{\rm 3D}_{xy} (0)= \frac{e^2}{h} \times (2n) \:\: \times \:\: \frac{{\rm K}_0}{2 \pi}
 \to \frac{n}{a} \:\: \frac{e^2}{h},
\end{equation}
when $\mu=0$, as shown in Fig.~\ref{Fig:AHE_Together}(a) for $n=1$, Fig.~\ref{Fig:AHE_Together}(b) for $n=2$ and Fig.~\ref{Fig:AHE_Together}(c) for $n=3$, since in our lattice realization of triple-component semimetals ${\rm K}_0=\pi/a$, where $a$ is the lattice spacing (see Sec.~\ref{Sec:LatticeModel}). Respectively $\sigma^{\rm 2D}_{xy}$ and $\sigma^{\rm 3D}_{xy}$ have dimension $\Omega^{-1}$ and $\Omega^{-1} \; {\rm m}^{-1}$. With increasing or decreasing chemical doping, the AHC decreases monotonically, as the contributions of the Berry curvature from the conduction and valence bands cancel each other. Scaling of the AHC as a function of chemical doping $\mu$ for three-dimensional triple-component semimetals is displayed in Fig.~\ref{Fig:AHE_Together} (lower panel). Finally note that the AHC in three-dimensions is insensitive to the presence of the trivial flat band. It should also be noted that in a gapless system (such as three-dimensional Weyl or triple-component semimetals) the AHC can be large, but generically not quantized, in contrast to the situation in an insulator, compare the upper and lower panels of Fig.~\ref{Fig:AHE_Together}. Also note that the AHC in such gapless topological systems is proportional to the separation of two triple-component points (K$_0$), see Eq.~(\ref{Eq:AHE_3D_results}), which is generically nonzero. Next we investigate the influence of the nontrivial Berry curvature in the medium on magnetotransport, such as longitudinal magnetoconductivity.


\section{Semiclassical Boltzmann transport}~\label{Sec:semiclassical}

In this section, we investigate the imprint of the nontrivial Berry curvature of a triple-component semimetal on various transport quantities within the framework of the semiclassical kinetic theory. Specifically, we compute (a) longitudinal magnetoconductivity [see Sec.~\ref{SubSec:LMC}], (b) planar Hall conductivity [see Sec.~\ref{SubSec:PHC}], and (c) longitudinal magneto-thermal conductivity [see Sec.~\ref{SubSec:LMTC}]. We also note that the nontrivial Berry curvature of integer pseudospin fermionic system can also leave its signature on chiral vortical effect~\cite{Sadofyev-1,Sadofyev-2}. In what follows we compute these quantities in the \emph{weak} magnetic field ($B$) limit, such that $\omega_c \tau \ll 1$, where $\omega_c$ is the cyclotron frequency and $\tau$ is the average time between two successive collisions (it should not be confused with the valley index $\tau=\pm$, introduced in Sec.~\ref{Sec:spin1_General}). In this limit one can neglect the Landau quantization, and treat $\tau$ to be independent of the strength of the external magnetic field. This approximation is justified since the radius of the cyclotron orbit in the weak field limit is sufficiently \emph{large}, allowing us to treat the path between two successive collisions as a \emph{straight line} (approximately). Furthermore, we also assume that there exists a single scattering life-time in the medium, determined by the elastic scattering of triple-component fermions from the impurities. On the other hand, for a sufficiently strong magnetic field ($\omega_c \tau \gg 1$) the Landau levels are sharply formed and one needs to account for the quantum corrections to $\tau$ due to the $B$-field, as the path between two successive collisions can no longer be treated as a straight line, and hence $\tau \equiv \tau(B)$.

In the presence of an external electric field ($\mathbf{E}$) and a temperature gradient ($\mathbf{\nabla T}$), the charge current ($\mathbf{J}$) and thermal current ($\mathbf{Q}$) are related to each other via the linear response equations, compactly written as
\begin{eqnarray}~\label{Eq:Current_Relations_Matrix}
\begin{pmatrix} 
\begin{array}{c} \mathbf{J} \\  \mathbf{Q}\end{array} \end{pmatrix} = \begin{pmatrix} \begin{array} {cc} \hat{\sigma} &  \hat{\alpha} \\ \hat{\bar{\alpha}} & \hat{\ell} \end{array} \end{pmatrix} \begin{pmatrix} \begin{array} {c} \mathbf{E} \\ -\mathbf{\nabla T}
\end{array}  
\end{pmatrix}
\end{eqnarray}
where $\hat{\sigma}$ is the conductivity tensor, $\hat{\alpha}$ is the Seebeck coefficient tensor, and $\hat{\ell}$ is the thermal conductivity tensor. Also note that $\hat{\bar{\alpha}}$ and $\hat{\alpha}$ are related to each other by Onsager's relation $\hat{\bar{\alpha}}=T\hat{\alpha}$. Within the framework of linear response theory the electrical and thermal currents can respectively be written as
\begin{eqnarray}
J_{a} &=& \sigma_{ab} \: E_{b} + \alpha_{ab}\: (-\nabla_{b} T)~\label{Eq:Electric_Current}, \\
Q_{a} &=& T\; \alpha_{ab} \: E_{b} + l_{ab} \: (-\nabla_{b} T)~\label{Eq:Thermal_Current},
\end{eqnarray}
where $a$ and $b$ are the spatial indices. We now set up the general formalism for the Boltzmann transport equations to compute these quantities in the presence of an underlying Berry curvature in the medium.

The Boltzmann transport equation reads as~\cite{Ziman}
\begin{equation}~\label{Eq:Boltzman}
\left(\partial_{t}+\mathbf{\dot{r}_{m}}\cdot\mathbf{\nabla_{r_{m}}}+\mathbf{\dot{k}_{m}}\cdot\mathbf{\nabla_{k_{m}}}\right)f_{m,\mathbf{k},\mathbf{r},t}=C \{f_{m,\mathbf{k},\mathbf{r},t}\},
\end{equation}
where $m$ is the band index and $C \{f_{m,\mathbf{k},\mathbf{r},t}\}$ is the collision integral, which in principle incorporates electron correlations (inelastic scattering) as well as elastic scattering from impurities, and $f_{m,\mathbf{k},\mathbf{r},t}$ is the electronic distribution function. For the sake of simplicity, we here focus only on the impurity scattering, which is the dominant source of relaxation process in weakly correlated and dirty systems. Within the relaxation time approximation, the collision integral takes the form 
\begin{equation}~\label{Eq:CollisionIntegral}
C\{f_{m,\mathbf{k}, \mathbf{r},t}\}=\frac{f_{m}^{0}-f_{m,k}}{\tau_{m}(\mathbf{k})}, 
\end{equation}
where $\tau_{m} (\mathbf{k})$ is the relaxation time and $f^{0}$ is the equilibrium Fermi-Dirac distribution function in the absence of any external field. To proceed further with the analysis, we ignore the momentum and band dependence of $\tau$ and assume it to be a constant with $\tau_{m} (\mathbf{k})=\tau$ (a phenomenological parameter) in the semiclassical limit, the \emph{single scattering-time approximation}.

\begin{figure}[t!]
\includegraphics[width=.95\linewidth]{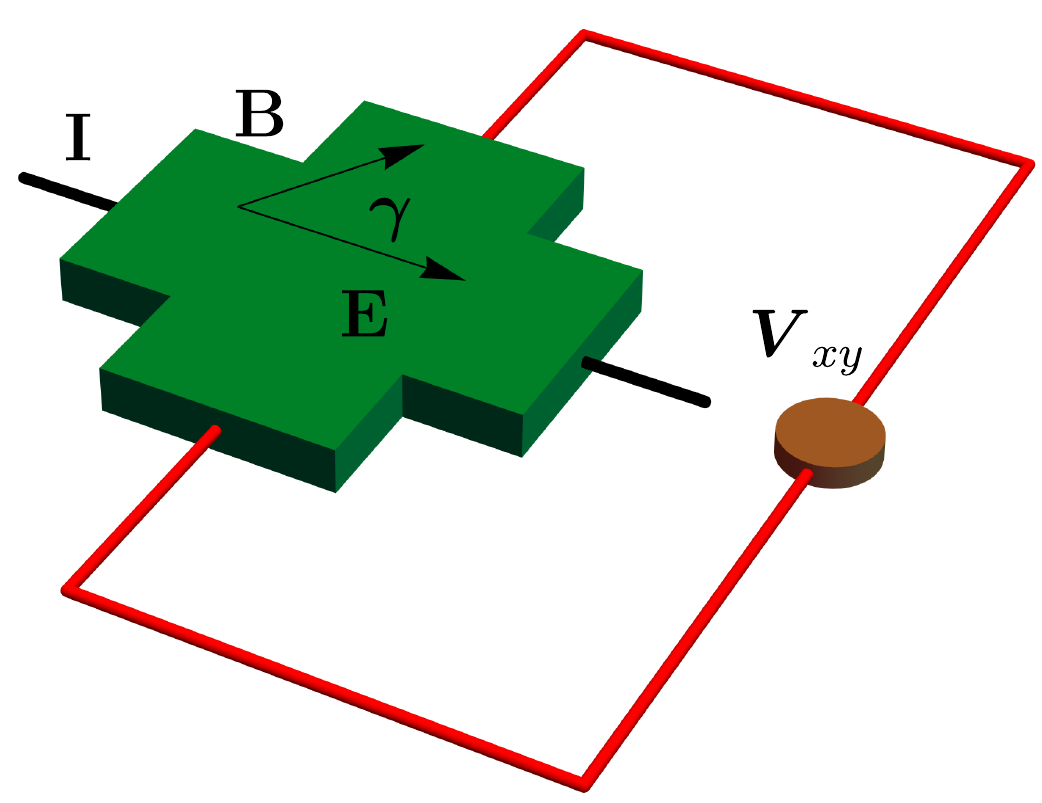}
\caption{A schematic setup for the measurement of the planar Hall conductivity. The electric field ($\mathbf{E}$) is applied along the $x-$axis and the magnetic field ($\mathbf{B}$) is confined within in the $xy$ plane. The angle $\gamma$ between ${\bf E}$ and ${\bf B}$ is measured with respect to the $x$-axis. See Sec.~\ref{SubSec:PHC} for the discussion on the planar Hall effect.}
\label{Fig:PHC_Setup}
\end{figure}

Upon incorporating the effects of the Berry curvature, the semiclassical equations of motion take the following form~\cite{Niu_1999,Niu_2006}
\begin{eqnarray}
\mathbf{\dot{r}_{m}} &=& \frac{1}{\hbar}\;\nabla\epsilon_{\mathbf{k}_{m}}- \left(\mathbf{\dot{k}_{m}}\times\mathbf{\Omega_{m,k}} \right),~\label{Eq:position_SC} \\
\mathbf{\dot{k}_{m}} &=& -\frac{e}{\hbar} \;\mathbf{E}-\frac{e}{\hbar} \; \left(\mathbf{\dot{r}_{m}} \times \mathbf{B} \right),~\label{Eq:momentum_SC}
\end{eqnarray}
where the second term of the Eq.~(\ref{Eq:position_SC}) represents the anomalous velocity originating from the nontrivial Berry curvature. The solutions of the coupled equations for $\mathbf{\dot{r}_{m}}$ and $\mathbf{\dot{k}_{m}}$ are respectively given by~\cite{Son_2012,Duval:2006}
\begin{eqnarray}
\mathbf{\dot{r}_{m}} &=& \frac{1}{D_{m}}[\mathbf{v_{m,k}}+\frac{e}{\hbar}(\mathbf{E}\times\mathbf{\Omega_{m,k}})+\frac{e}{\hbar}(\mathbf{v_{m,k}}\cdot\mathbf{\Omega_{m,k}})\mathbf{B}],~\label{Eq:position_solution} \nonumber \\
\\
\mathbf{\dot{k}_{m}} &=& \frac{1}{\hbar D_{m}}[e\mathbf{E}+\frac{e}{\hbar}(\mathbf{v_{m,k}} \times \mathbf{B})+\frac{e^{2}}{\hbar}(\mathbf{E}\cdot\mathbf{B})\mathbf{\Omega_{m,k}}].~\label{Eq:momentum_solution}
\end{eqnarray}
For brevity we use $D_m \equiv D_{m}(\mathbf{B,\Omega_{m,k}})$ in the above two equations, where 
\begin{equation}
D_{m}(\mathbf{B,\Omega_{m,k}})=\left[ 1+\frac{e}{\hbar}\: \left( \mathbf{B} \cdot \mathbf{\Omega_{m,k}} \right) \right] \nonumber 
\end{equation}
modifies the invariant phase space volume according to $dkdx \rightarrow D_{m}(\mathbf{B,\Omega_{m,k}})dkdx$ and gives rise to a noncommutative mechanical model, since the Poisson bracket of two coordinates is now \emph{nonzero}~\cite{Duval:2006}. We are now equipped to proceed to the computation of various conductivity tensors introduced in Eqs.~(\ref{Eq:Current_Relations_Matrix})-(\ref{Eq:Thermal_Current}). The rest of the analysis is presented only for linear-triple component semimetals (with $n=1$). The following discussion can be generalized to address similar effects in quadratic and cubic-triple component semimetals, and establish the scaling of various components of the conductivity tensor with the monopole charge $2n$. We leave this exercise for a future investigation.

\begin{figure*}[t!]
\subfigure[]{
\includegraphics[width=.32\linewidth]{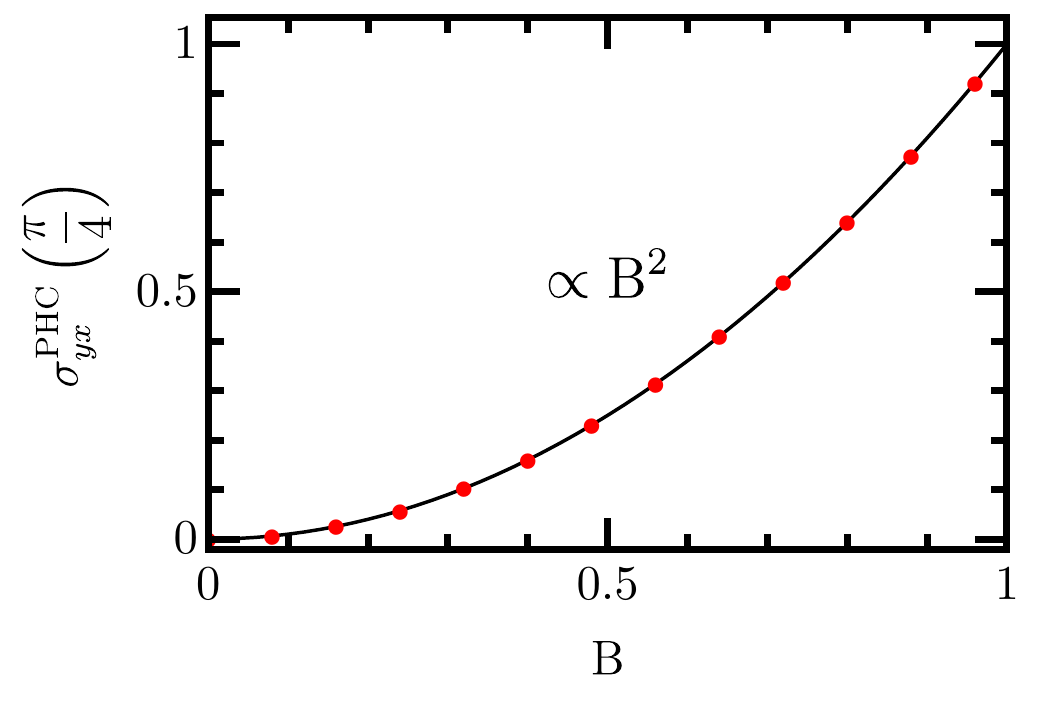}
}%
\subfigure[]{
\includegraphics[width=.32\linewidth]{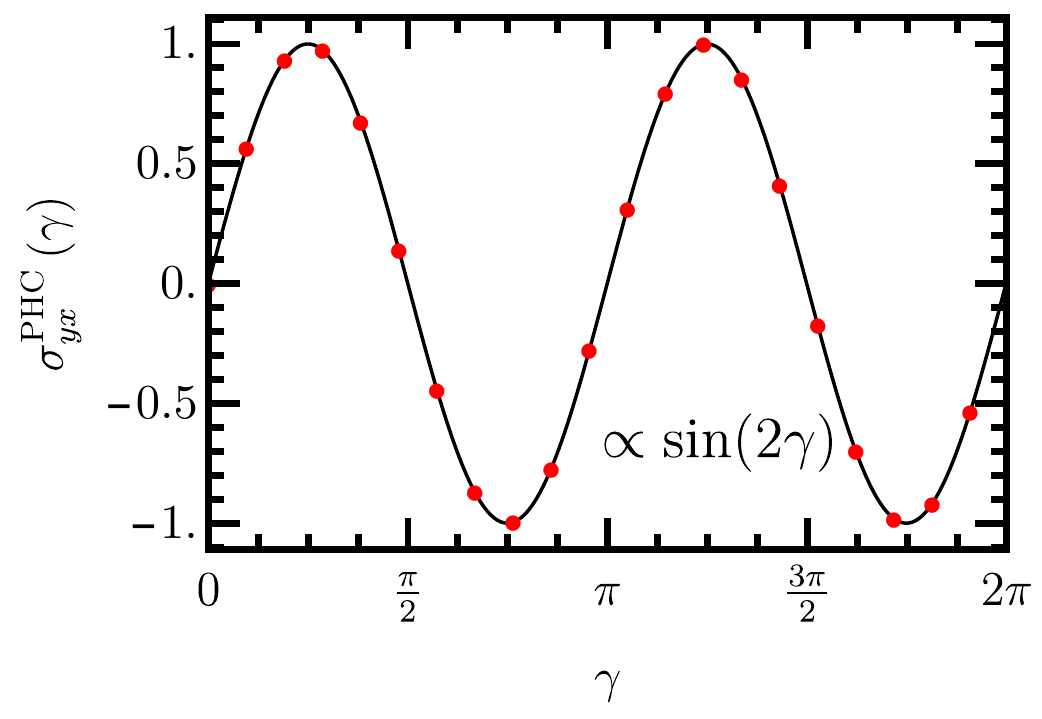}
}%
\subfigure[]{
\includegraphics[width=.32\linewidth]{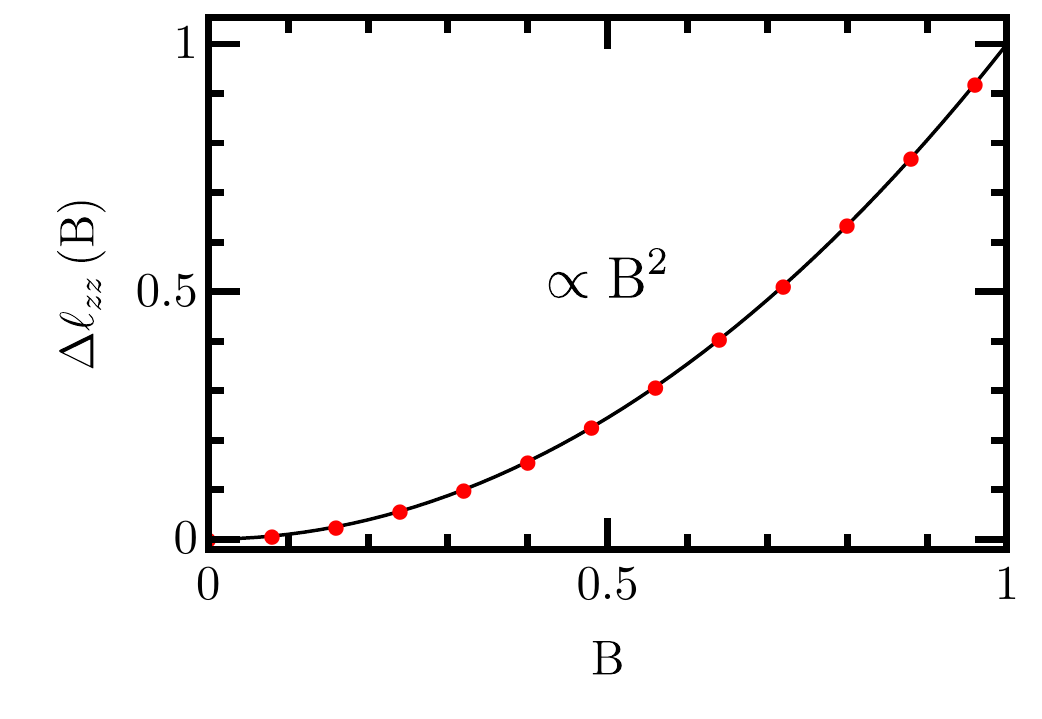}
}%
\caption{(a) The amplitude of the planar Hall conductivity (normalized by its maximal value for $B=B_{\rm max}$) as a function of the magnetic-field strength $B$ (normalized by $B_{\rm max}$) in a linear triple-component semimetal for $\gamma=\frac{\pi}{4}$, see Eq.~(\ref{Eq:PHE_EBConfiguration}). The amplitude of the planar Hall conductivity scales as $B^2$, see Sec.~\ref{SubSec:PHC}. (b) The angular dependence of the planar Hall conductivity (normalized by its amplitude for $\gamma=\frac{\pi}{4}$), showing a $\sin (2\gamma)$ scaling. (c) The scaling of the longitudinal magnetothermal conductivity $\Delta \ell_{zz} (B)=\ell_{zz}(B)-\ell_{zz}(0)$ (normalized by its maximal value at $B=B_{\rm max}$) as a function of $B$ (normalized by $B_{\rm max}$). Note that ${\bf B}$ and $\mathbf{\nabla T}$ is along the $z$ direction, and $\Delta \ell_{zz} (B) \sim B^2$, see Sec.~\ref{SubSec:LMTC} for details. Numerical calculations of $\sigma^{\rm PHC}_{yx} (\gamma)$ and $\Delta \ell_{zz}(B)$ are performed from the low-energy model for the linear triple-component semimetal [see Sec.~\ref{Sec:spin1_General}]. The red dots represent numerically computed values of planar Hall [panels (a) and (b)] and magnetothermal [panel (c)] conductivitities, whereas the black curves show a fitting with $B^2$ [panels (a) and (c)] and $\sin(2\gamma)$ [panel (b)]. 
}
\label{Fig:PHC_results}
\end{figure*}

     \subsection{Longitudinal magnetoconductivity}~\label{SubSec:LMC}

For the computation of the longitudinal magnetoconductivity (LMC), we assume that $\mathbf{E}$ and $\mathbf{B}$ are always parallel to each other, otherwise applied along an arbitrary direction $\hat{\mathbf{u}}$. After solving the Boltzmann equation using Eqs.~(\ref{Eq:position_solution}) and ~(\ref{Eq:momentum_solution}), we find~\cite{Son:2013,Kim:2014,Lundgren:2014,Sharma:2016}
\begin{eqnarray}~\label{Eq:LMC_first}
\sigma_{uu} (B) &=& -e^{2}\tau\sum_{m}\int \frac{d^{3}{\bf k}}{(2\pi)^{3}} \:\: \left[ 1+\frac{e}{\hbar}\: \left( \mathbf{B} \cdot \mathbf{\Omega_{m,k}} \right) \right]^{-1}   \nonumber \\
&\times& \left[ {v_{m,u}} +\frac{eB}{\hbar}(\mathbf{v_{m,k}}\cdot\mathbf{\Omega_{m,k}}) \right]^{2} \:\: \left( \partial_{\epsilon}f_{m}^{0} \right).
\end{eqnarray}
For concreteness, we compute the LMC in the $z$ direction ($\sigma_{zz}$) from the linearized model, introduced in Sec.~\ref{Sec:spin1_General}. Note that only the two dispersive bands contribute to the LMC, as the carriers in flat band are \emph{localized}. As temperature $T \rightarrow 0$, the above expression for the LMC simplifies to (after setting $\hbar=1$)  
\begin{align}~\label{Eq:LMC_T0}
\sigma_{zz} (B)=e^{2}\tau\int \frac{d^3\mathbf{k}}{(2\pi)^3}\frac{\left[ v_z+eB \left( \mathbf{v_{k}}\cdot\mathbf{\Omega_{k}} \right) \right]^{2}}{1+eB\Omega_{z}} \: \delta(\mu-\epsilon_{\mathbf{k}}),
\end{align}
where we have used the fact that $\lim_{T \to 0}\partial_{\epsilon}f^{0}=-\delta(\mu-\epsilon_{\mathbf{k}})$. For rest of the analysis, we set $\mu > 0$, so that only the upper band contributes to the LMC. Upon introducing the polar coordinates in which ${\bf k}=k\left(\cos \phi \sin \theta, \sin \phi \sin \theta, \cos \theta \right)$, Eq.~(\ref{Eq:LMC_T0}) can be written as
\begin{widetext}
\begin{align}
\sigma_{zz} (B)&=\frac{e^{2}\tau}{(2\pi)^2} \int_{0}^{\pi} d\theta \sin \theta  \int_{0}^{\infty}dk k^2 \frac{(\cos \theta + \frac{eB}{k^2})^{2}}{1+\frac{eB\cos \theta}{k^2}}\delta(\mu-k) 
=\frac{e^{2}\tau \mu^2}{(2\pi)^2} \Bigg[ \int_{0}^{\pi} d \theta \; \sin \theta \cos^2 \theta \left( 1+\frac{eB\cos \theta}{\mu^2} \right)^{-1} \nonumber \\
&+ 2 \frac{eB}{\mu^2} \int_{0}^{\pi} d\theta \; \sin \theta \cos \theta  \left( 1+\frac{eB\cos \theta}{\mu^2} \right)^{-1}  + \frac{e^2 B^2}{\mu^4} \int_{0}^{\pi} d\theta\;  \sin \theta  \left( 1+\frac{eB\cos \theta}{\mu^2} \right)^{-1}  \Bigg] \nonumber \\
&= \frac{e^{2}\tau \mu^2}{(2\pi)^2} \Bigg[  -\frac{2 \mu^4}{e^2 B^2}   +   \frac{2 \mu^6}{e^3 B^3} \tanh^{-1}  \left(\frac{eB}{\mu^2} \right) +   \frac{2eB}{\mu^2} \left\{ \frac{2 \mu^2}{e B} + \left( 1- \frac{2 \mu^4}{e^2 B^2} \right) \tanh^{-1}  \left( \frac{eB}{\mu^2} \right)   \right\}  \Bigg].
\end{align}
\end{widetext}
While arriving at the last expression we set $v_{z}=1$. Note that semiclassical theory is applicable in the parameter regime where quantum corrections can be neglected. Such condition at $T=0$ is achieved only if $\sqrt{e B} \ll \mu$, so that the chemical potential provides the infrared cut-off in the system. Therefore, we can expand $\tanh^{-1}(x)$ appearing in the above expression for $x \ll 1$, yielding 
\begin{equation}
\tanh^{-1} (x) = x - \frac{x^3}{3} + {\mathcal O}\left( x^5 \right),
\end{equation}
where $x=eB/\mu^2$. Finally, accounting for the contributions from two triple-component nodes, we find 
\begin{align}~\label{Eq:LMC_Final}
\sigma_{zz} (B)= 2 \times \left( \frac{e^{2}\tau \mu^2}{4 \pi^2} \right) \: \Big[    \frac{2}{3} + \frac{16}{15}  \frac{e^2 B^2}{\mu^4}  \Big].
\end{align}
The first term in the above expression is the standard metallic conductivity arising from the Drude contribution, while the second term shows a $B^2$ enhancement of the LMC. However, in the single scattering time approximation we cannot attribute such an enhancement solely to the chiral anomaly, as both the regular and axial charges are relaxed by the same scattering mechanism (characterized by $\tau$)~\cite{roy-surowka}. Nonetheless, the system still displays a positive LMC and $\sigma_{zz} \sim B^2$. In order to isolate the contribution from the chiral anomaly we need to introduce two scattering times in the collision integral [see Eq.~(\ref{Eq:CollisionIntegral})], $\tau_{\rm inter}$ and $\tau_{\rm intra}$, respectively denoting the inter- and intravalley scattering lifetimes. In particular, when $\tau_{\rm inter} \gg \tau_{\rm intra}$ only the contribution from the chiral anomaly survives~\cite{zyuzin}. Explicit demonstration for this lengthy analysis is left for a future investigation.  

  	 \subsection{Planar Hall Conductivity}~\label{SubSec:PHC}

The planar Hall effect corresponds to the appearance of an in-plane transverse voltage ($V_{xy}$) in the presence of external, but \emph{coplanar} electric and magnetic fields; specifically when they are not perfectly aligned to each other. The experimental setup for the measurement of planar Hall conductivity (PHC) is schematically shown in Fig.~\ref{Fig:PHC_Setup}. Notice that in this configuration the conventional Hall effect vanishes. To evaluate the PHC, we conveniently align the electric field ($\mathbf{E}$) along the $x-$axis, while the magnetic field ($\mathbf{B}$) is directed at a finite angle $\gamma$ from the $x$-axis (but in the $xy$ plane), thus 
\begin{equation}~\label{Eq:PHE_EBConfiguration}
\mathbf{E}=E\;\hat{x}, \quad {\rm and} \quad\mathbf{B}=B \left(\cos\gamma \; \hat{x}+\sin\gamma \;\hat{y}\right), 
\end{equation}
where $\gamma$ is the angle between $\mathbf{E}$ and $\mathbf{B}$ (see Fig.~\ref{Fig:PHC_Setup}). The PHC is then given by~\cite{Burkov_2017,Nandy_2017,Nandy1_2017}
\begin{widetext}
\begin{eqnarray}~\label{Eq:PHC_Definition}
\sigma^{\rm PHC}_{yx} (\gamma) = -e^{2} \tau \sum_{m}\int \frac{d^3{\bf k}}{(2 \pi)^3} \: \frac{v_{m,y}+\frac{eB \sin \gamma}{\hbar} \left( \mathbf{v_{m,k}}\cdot\mathbf{\Omega_{m,k}} \right)}{ 1+\frac{e}{\hbar}\: \left( \mathbf{B} \cdot \mathbf{\Omega_{m,k}} \right)} \: 
\left[v_{m,x}+\frac{eB \cos \gamma}{\hbar} \left( \mathbf{v_{m,k}}\cdot\mathbf{\Omega_{m,k}} \right) \right] \left(\frac{\partial f^{0}}{\partial \epsilon}\right).
\end{eqnarray}
As $T \to 0$ in terms of the polar coordinates, introduced earlier, the PHC reads as (after setting $\hbar=1$) 
\begin{align}
	\sigma_{yx}(\gamma) =\frac{e^{2}\tau \mu^2}{(2\pi)^3} \int_{0}^{2\pi} d\phi \int_{0}^{\pi} d\theta \left[1+\frac{eB\sin \theta \cos (\phi-\gamma)}{\mu^2}\right]^{-1} \left[ \sin^3 \theta \frac{\sin 2\phi}{2} + \frac{eB}{\mu^2}\sin^2 \theta \sin (\phi+\gamma)+ \frac{eB}{2\mu^4}\sin \theta \sin 2\gamma \right].
\end{align}
\end{widetext}
We numerically compute the PHC from the low-energy model for a linear triple-component semimetal. The amplitude of the PHC shows a $B^{2}$ dependence, as shown in Fig.~\ref{Fig:PHC_results}(a), for any value of $\gamma$ except when $\gamma=0$ and $\gamma=\frac{\pi}{2}$, where PHC vanishes. Also note that the PHC scales as $\sin (2\gamma)$, see Fig.~\ref{Fig:PHC_results}(b). We find that the PHC for TCF does not satisfy the antisymmetry property ($\sigma_{xy}=-\sigma_{yx}$) of the regular Hall conductivity since it does not originate from the conventional Lorentz force.

		 \subsection{Longitudinal Magneto-thermal Conductivity}~\label{SubSec:LMTC}

Next we compute the longitudinal magneto-thermal conductivity (LMTC) for the linear triple-component semimetal from its low-energy model. To compute the LMTC, we align the external magnetic field $\mathbf{B}$ and the temperature gradient $\mathbf{\nabla T}$ along an arbitrary direction $\hat{\mathbf{u}}$, such that ${\bf B} \parallel \mathbf{\nabla T}$. After solving the Boltzmann equation using Eqs.~(\ref{Eq:position_solution}) and~(\ref{Eq:momentum_solution}), and comparing with Eq.~(\ref{Eq:Thermal_Current}), we arrive at the following expression for the LMTC~\cite{Lundgren:2014, Sharma:2016, Nandy2_2017} 
\allowdisplaybreaks[4]
\begin{eqnarray}~\label{Eq:LMTC_definition}
\ell_{uu}&=&\tau \sum_{m}\int \frac{d^3{\bf k}}{(2\pi)^3} \; \left[ 1+\frac{e}{\hbar}\: \left( \mathbf{B} \cdot \mathbf{\Omega_{m,k}} \right) \right]^{-1} \\
&\times& \left[ v_{m,u} + \frac{eB}{\hbar} (\mathbf{v_{m,k}}\cdot\mathbf{\Omega_{m,k}})\right]^{2} 
\frac{\left( \epsilon_{m}-\mu \right)^2}{T}\left( -\partial_{\epsilon}f_{m}^{0} \right). \nonumber 
\end{eqnarray}
Since the flat band is topologically trivial (possessing zero Chern number), and the chemical potential is placed above the triple-component points, only upper band contributes to the LMTC. We compute LMTC separately for each triple-component node and finally add their contributions. For concreteness, we compute the LMTC along the $z$ direction. In terms of the polar coordinates and after setting $v_z=1$, $\hbar=1$, we arrive at the final expression for the LMTC at finite-$T$, given by  
\begin{align}~\label{Eq:LMTC_Final}
\ell_{zz}&=\frac{e^{2}\tau}{(2\pi)^2} \int_{0}^{\pi}\sin \theta d\theta \int_{0}^{\infty}dk \frac{(k-\mu)^2}{T^2} k^2 \nonumber \\
& \times \frac{(\cos \theta + \frac{eB}{k^2})^{2}}{1+\frac{eB\cos \theta}{k^2}} \: f^{0}(1-f^{0}),
\end{align}
where we have used the fact that $-\partial_{\epsilon}f^{0}=f^{0}\left( 1-f^{0} \right) /T$. The scaling of the LMTC, specifically $\Delta\ell_{zz} (B)=\ell_{zz}(B)-\ell_{zz}(0)$, as a function of the magnetic-field strength ($B$) is displayed in Fig.~\ref{Fig:PHC_results}(c). We find that LMTC also scales as $B^2$, i.e. $\Delta\ell_{zz} (B) \sim B^{2}$ for linear triple-component semimetals.


\section{Conclusions and Discussions}~\label{Sec:summary}

To summarize, we generalize the notion of time-reversal symmetry-breaking pseudospin-1 or triple component semimetals to arbitrary integer (anti-)monopole charge $2 n$ [see Secs.~\ref{Sec:spin1_General} and \ref{Sec:LatticeModel}] and address its topological properties, such as the Fermi arc surface states [see Secs.~\ref{Sec:SurfaceStates}]. In addition, we also compute the influence of the nontrivial Berry curvature in this system on various transport quantities, such as the anomalous Hall conductivity [see Sec.~\ref{Sec:AHE}] within the framework of the Kubo formalism, as well as the longitudinal magnetotransport and the planar Hall conductivity using the semiclassical Boltzmann theory in the single scattering time approximation [see Sec.~\ref{Sec:semiclassical}].

In particular, we show that on a simple cubic lattice one can realize triple component nodes with monopole charge $2n$, where $n=1,2,3$ in a crystalline environment [see Sec.~\ref{Sec:LatticeModel}]. At the triple-component points three bands with pseudospin quantum numbers $m_s=1,0,-1$ touch each other. While two bands with pseudospin projections $|m_s|=1$ are dispersive away from the triple-component points, the one with $m_s=0$ is completely flat and topologically trivial. In our lattice realization of the spin-1 fermions, the triple-component points are separated along the $z$-direction. For any $n$, the energy dispersion (for $|m_s|=1$ bands) always scales linearly with $k_z$, but $E_{{\bf k}_\perp} \sim k^n_\perp$, where $k_\perp=\left[ k^2_x + k^2_y\right]^{1/2}$. We also show that such unusual band touchings are invariant under discrete four-fold or $C_4$ rotations and can be realized from simple tight-binding models in cubic lattice [see Sec.~\ref{Sec:LatticeModel}].

The topological invariant for triple-component points manifests through the Fermi arc surface states, following the bulk-boundary correspondence. We argue that a system with a pair of triple-component points with (anti-)monopole charge $2 n$ accommodates $2 |n|$ branches of Fermi arc states on the surface, see Sec~\ref{Sec:SurfaceStates}. To establish the bulk-boundary correspondence for spin-1 triple-component fermions, we numerically diagonalize the tight-binding models for these systems [introduced in Sec.~\ref{Sec:LatticeModel}] with periodic boundary in the $y$ and $z$ directions (hence momenta along these two directions are good quantum numbers) and a open boundary in the $x$-direction. Figure~\ref{Fig:FermiArc_2D} depicts the Fermi arcs in the $\left( k_y,k_z\right)$ plane (the top surface), and we find that there exists exactly $2|n|$ branches of the Fermi arc surface states on the top surface connecting two triple-component points. Additional salient features of the arc states can be appreciated from their localization in the $x$-direction, as shown in Fig.~\ref{Fig:FermiArc_3D}. We find that while the Fermi arc states are well localized on the top or bottom surfaces away from the triple-component points, at these two points (representing singularities in the momentum space) they are completely \emph{delocalized}. Specifically, the arcs states from the top surface leak through the bulk triple-component points, and get connected to the ones on the bottom surface. This feature is insensitive to the precise value of $n$ and also occurs for spin-1/2 Weyl fermions.

The two dispersive bands in triple-component semimetals possess nontrivial Berry curvature, whereas the flat band is topologically trivial. The signature of nontrivial Berry curvature can, for example, be observed in the anomalous Hall conductivity [see Sec.~\ref{Sec:AHE}]. Note that time-reversal symmetry-breaking triple-component semimetals can be envisioned as stacking of two-dimensional anomalous Hall or Chern insulators of spin-1 fermions in the momentum space along $k_z$ direction between two triple-component nodes. As a result for $k_z=0$ we obtain a quantized anomalous Hall conductivity, given by $\sigma^{\rm 2D}_{xy}=2 n e^2/h$, for $n=1,2,3$, see Figs.~\ref{Fig:AHE_Together}(a)-(c), when the chemical potential lies within the bulk band gap. The anomalous Hall conductivity of a triple-component semimetal can then be obtained by accumulating the quantized contribution from each two-dimensional constituting layers in between two nodes, and the results are summarized in Figs.~\ref{Fig:AHE_Together}(d)-(f). Therefore, the generalization of spin-1 topological semimetals opens up a new route to achieve large anomalous Hall conductivity. A large (but not quantized) anomalous Hall conductivity can also be accommodated by spin-1/2 Weyl fermions, which can be germane for Pr$_2$Ir$_2$O$_7$ inside a metallic spin-ice ordered phase~\cite{goswami-roy-dassarma, andras-roy}. In particular, in Pr$_2$Ir$_2$O$_7$ the biquadratic touching of the Kramers degenerate valence and conduction bands, describing the normal state of 227 pyrochlore iridates~\cite{QBT_iridates1, QBT_iridates2}, can be destabilized by the onset of a spin-ice or three-in one-out magnetic ordering for itinerant fermions, which gives birth to only two Weyl nodes and concomitantly supports anomalous Hall conductivity $\sim 10^3 \; \Omega^{-1} {\rm m}^{-1}$~\cite{Pr2Ir2O7_AHE}. Therefore, possible material realizations of spin-1 Weyl fermions in strongly correlated systems should be an interesting future avenue of research.

The signature of the Berry curvature can also be found in various other transport quantities, such as longitudinal magneto- and magnetothermal conductivities, planar Hall conductivity [see Sec.~\ref{Sec:semiclassical}]. We here compute these quantities using the semiclassical Boltzmann transport theory and for sufficiently weak magnetic field, when the Landau levels are not sharply formed. At least when the strength of the external magnetic field ($B$) is sufficiently weak, all of them increase as $B^2$. Even though it is commonly believed that such a seemingly counter intuitive enhancement of the longitudinal magnetoconductivity, for example, arising from the nontrivial Berry curvature captures the signature of the chiral anomaly, there exists no concrete proof demonstrating this connection. Therefore, it will be interesting to investigate these quantities in the strong magnetic field limit when the inter-particle scattering time ($\tau$) explicitly depends on the magnetic field, and establish the relevance of chiral anomaly in spin-1 system~\cite{lepori, argyres-adams, li-roy-dassarma}.

\begin{figure}[t!]
	\includegraphics[width=0.45\textwidth]{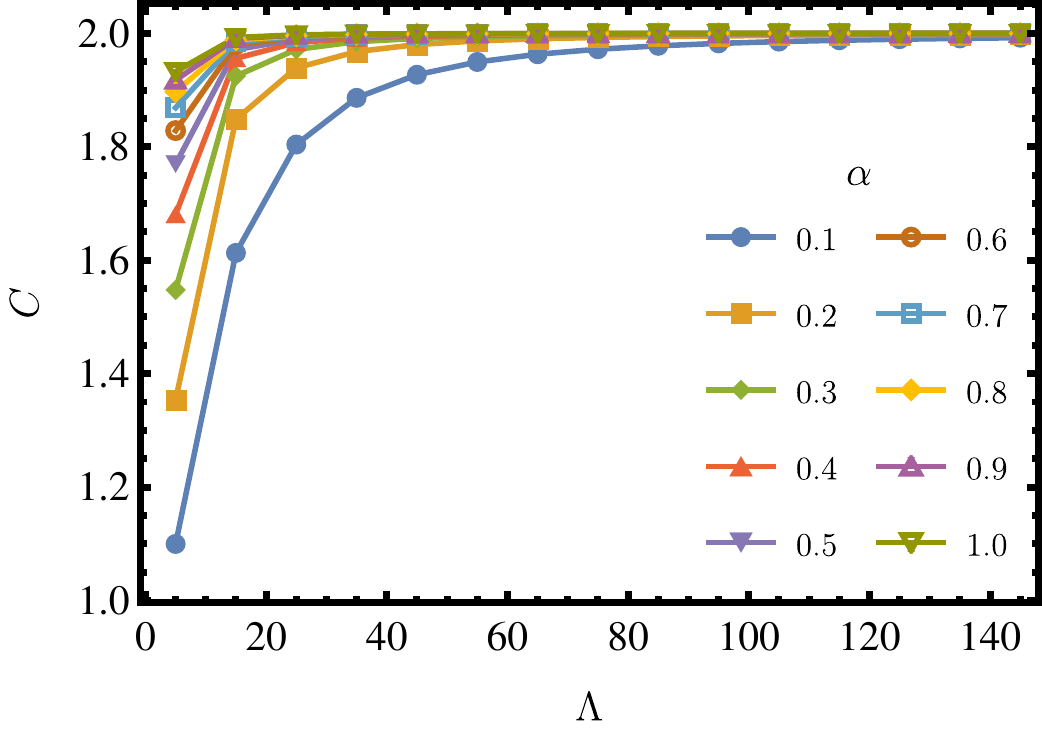}
	\caption{ 
	Numerically computed Chern number ($C$) of the upper band of a two-dimensional anomalous Hall insulator, occupying the $k_z=0$ plane of a linear triple-component semimetal, as a function of the ultraviolet momentum cut-off $\Lambda$ [see Eq.~(\ref{Eq:Ham_Chern})] for different values of $\alpha$ [see Eq.~(\ref{Eq:Chern})].	}~\label{alpha_c}
\end{figure}

Our discussion is, however, not limited to spin-1 or triple-component fermions. For example, our lattice construction for spin-1 fermions from Sec.~\ref{Sec:LatticeModel} can immediately be generalized to any integer spin-$s$ fermions by replacing the spin-1 matrices by spin-$s$ matrices. In that construction, there always exists a topological trivial flat band, and $2 s$ dispersive bands ($s$ number of valence and conduction bands), characterized by $s$ distinct Fermi velocities. Therefore, our theoretical analysis should stand as a good starting point to begin the voyage into the world of integer spin topological phases of matter. Besides the topological features of integer-spin Weyl systems, its (in)stability against electronic correlations, which can give birth to exotic superconducting~\cite{rahul-spin-1-SC} and excitonic phases, is yet another interesting avenue, which we will explore in future.


\acknowledgements

S. N. acknowledges MHRD, India for research fellowship. B. R. is thankful to Nordita for hospitality during the program ``Topological Matter Beyond the Ten-Fold Way", and Vladimir Juri\v ci\' c and Soumya Bera for valuable discussions. BR was partially supported by the start-up grant from Lehigh University. 


\appendix


\section{Ultraviolet regularization and Chern number}~\label{App:UVRegularization}

The AHC of two-dimensional constituting layers of the anomalous Hall insulators (AHI) is intimately tied with the first Chern number of the system in the following way
\begin{equation}
\sigma^{\rm 2D}_{xy} (\mu=0,k_z=0)=\frac{e^2}{h} \times \text{Chern number of the AHI}.
\end{equation} 
We established this connection by explicitly computing the AHC from tight-binding models of the generalized triple-component semimetals, introduced in Sec.~\ref{Sec:LatticeModel}, after setting $k_z=0$ (thus yielding an AHI), see also Figs.~\ref{Fig:AHE_Together}(a)-\ref{Fig:AHE_Together}(c). This Appendix is devoted to illustrate how the correct Chern number can be extracted from the continuum models of these systems.

 \begin{figure*}
\subfigure[]{
 	\includegraphics[width=0.45\textwidth]{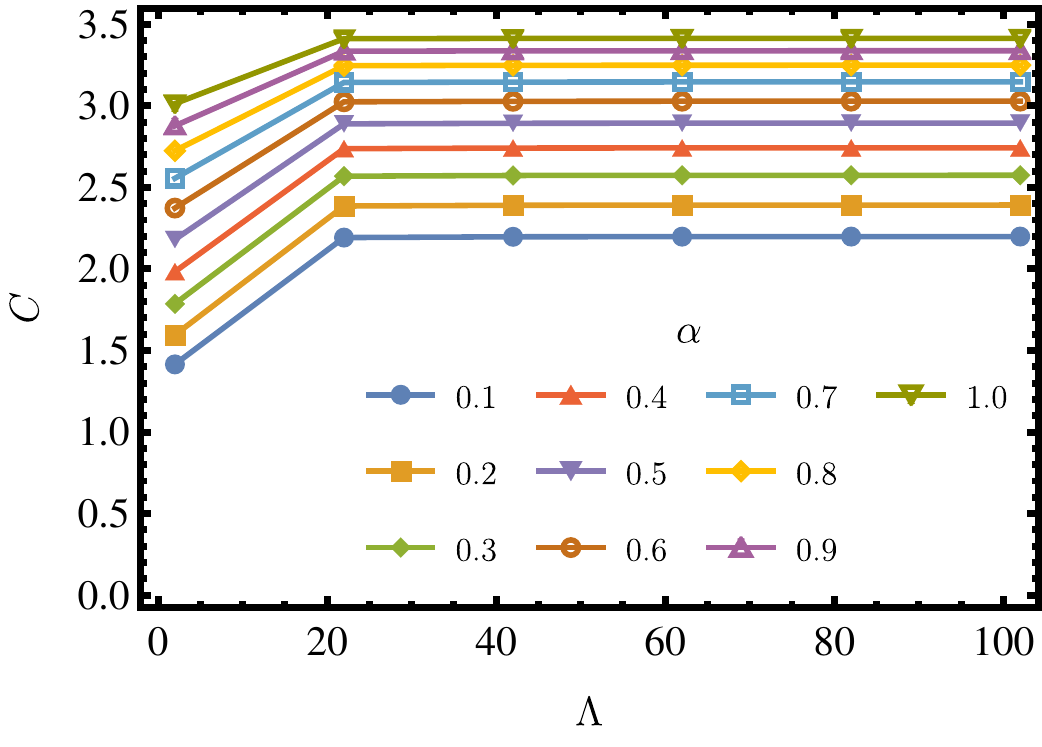}
	}%
\subfigure[]{
	\includegraphics[width=0.435\textwidth]{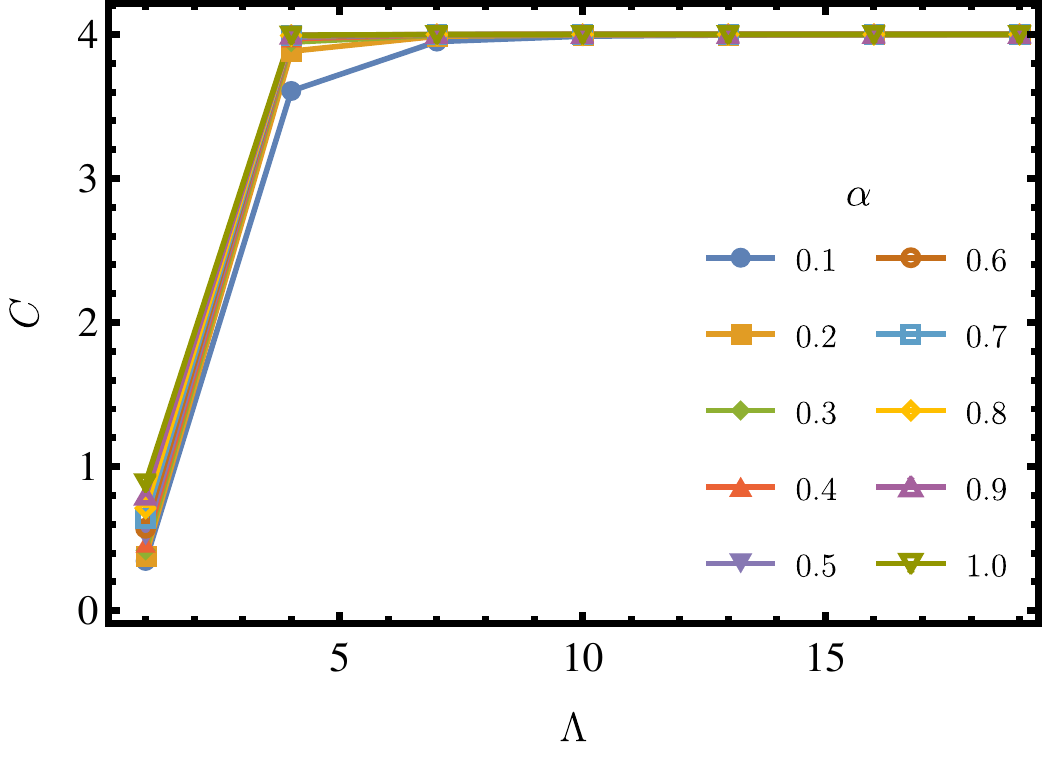}
	}
\subfigure[]{	
 	\includegraphics[width=0.45\textwidth]{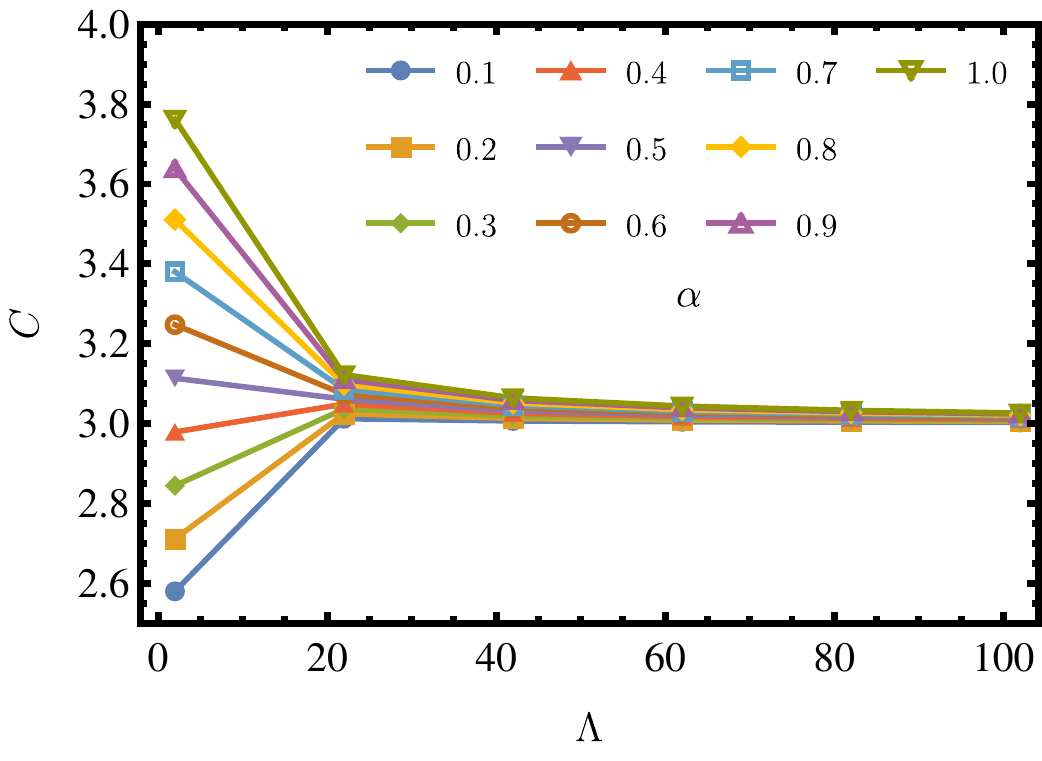}
	}%
\subfigure[]{		
 	\includegraphics[width=0.45\textwidth]{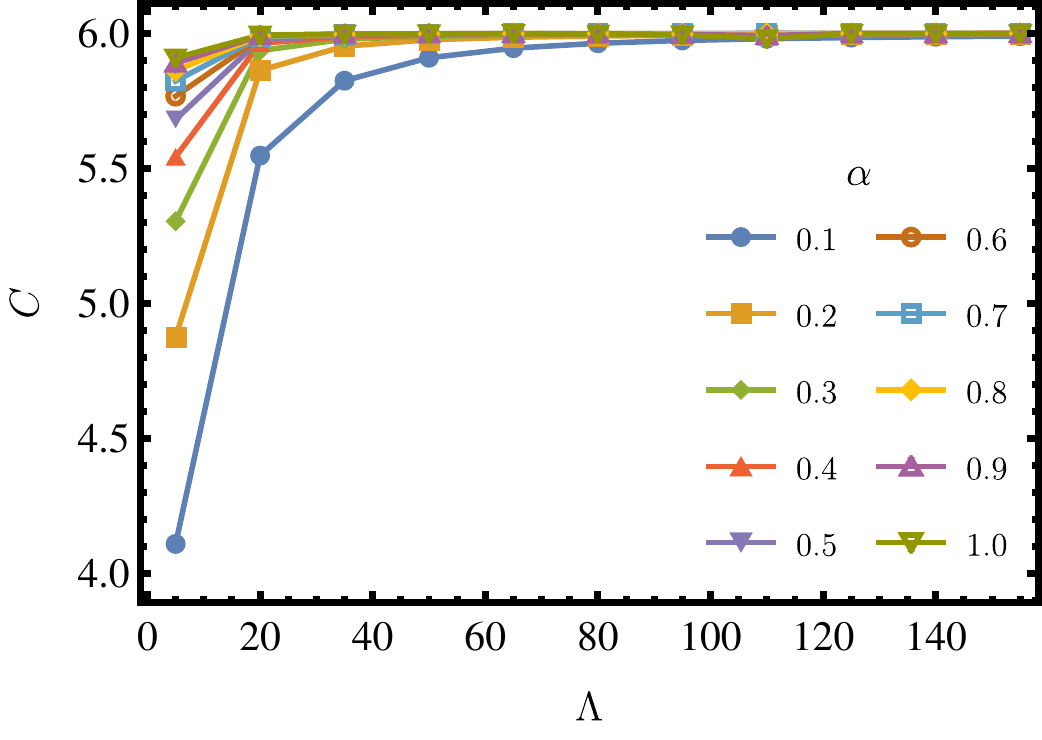}
	}
\caption{ Numerically computed Chern number ($C$) of the upper band of the two-dimensional anomalous Hall insulator (AHI), occupying the $k_z=0$ plane of the quadratic [panels (a) and (b)] and cubic [panels (c) and (d)] triple-component semimetals, as a function of the ultraviolet momentum cut-off $\Lambda$ [see Eq.~(\ref{Eq:Ham_Chern})] for different values of $\alpha$ [see Eq.~(\ref{Eq:TopoHamil_lowenergy})]. For panels (a) and (c), we take $f \left({\bf k}_\perp \right)=k^2_x+k^2_y$ [obtained from the leading-order expansion of $N^2_z({\bf k})$ introduced in Eq.~(\ref{Eq:LatticeModel_Components})] and the Chern numbers do not reveal the correct topological invariant of the AHI. By contrast, in panels (b) and (d) we compute the Chern number with $f \left({\bf k}_\perp \right)=k^4_x+k^4_y$ [obtained from the leading-order expansion of $N^2_z({\bf k})$ introduced in Eq.~(\ref{Eq:Wilsonmass_New})] and we recover the correct topological invariant of the AHI for sufficient large $\Lambda$. See Appendix~\ref{App:UVRegularization} for detailed discussion on this issue. }~\label{Chern_D_T}
 \end{figure*}

\begin{figure*}[t!]
\subfigure[]{
\includegraphics[height=5.00cm]{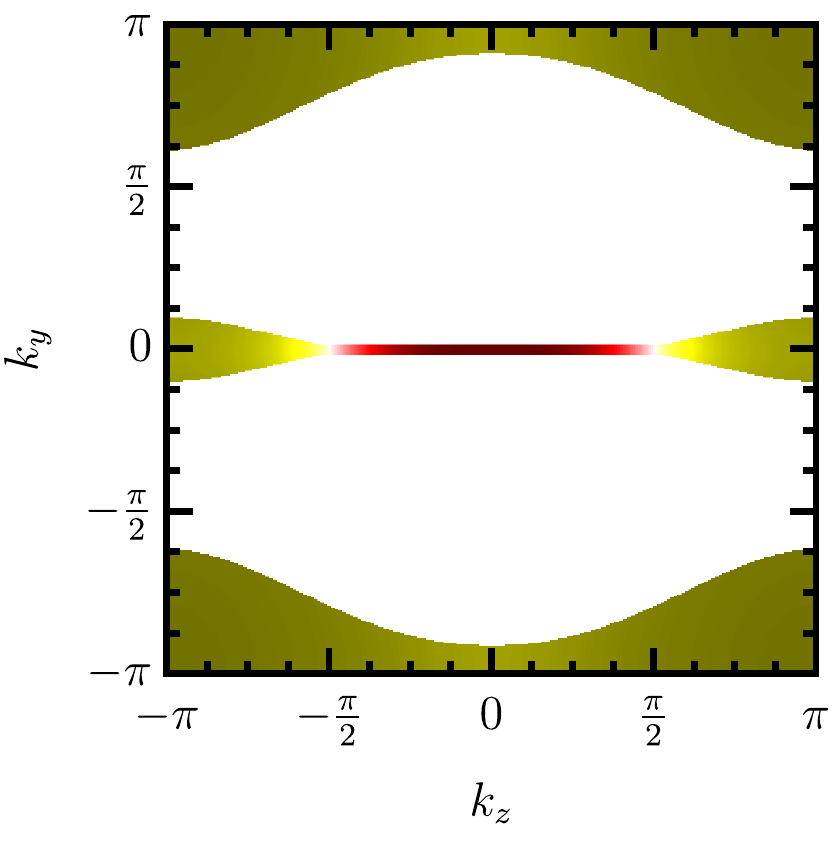}
}%
\subfigure[]{
\includegraphics[height=5.00cm]{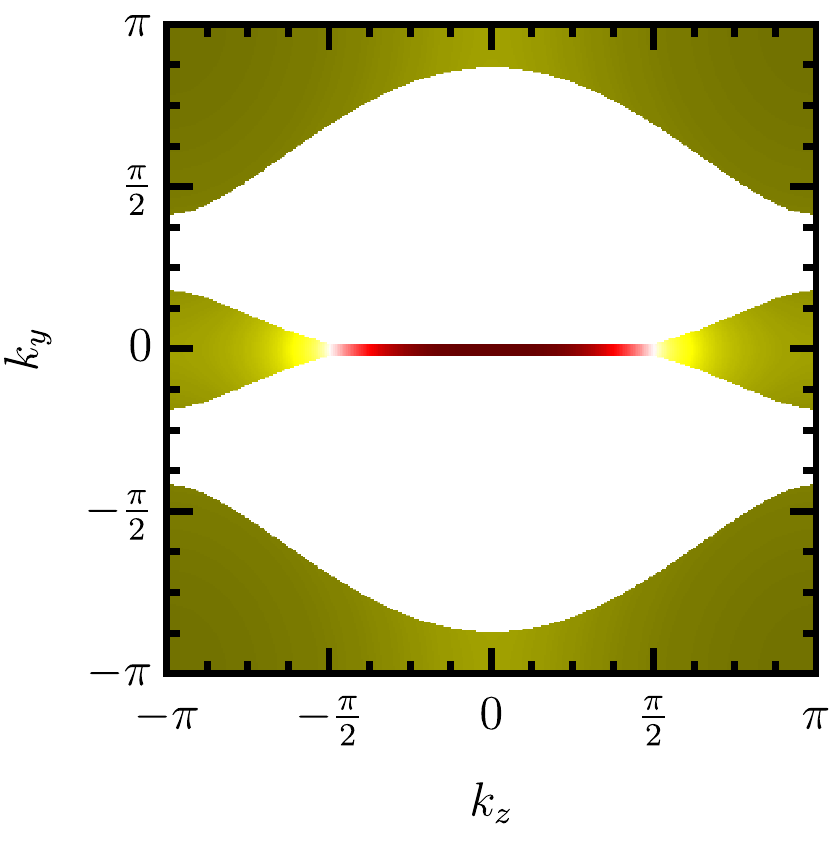}
}%
\subfigure[]{
\includegraphics[height=5.00cm]{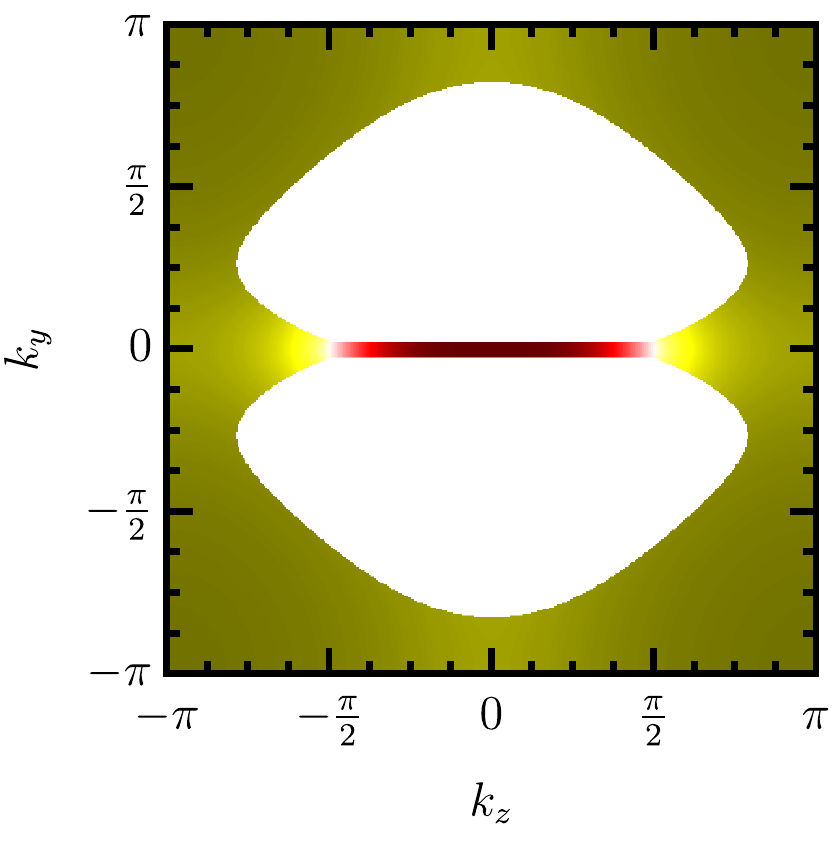}
}%
\includegraphics[height=5.00cm]{ColorBar_color_6.pdf}%
\caption{The amplitude of the surface localized states residing on the top surface and within the energy window $\Delta E= 0.04 t$ for (a), $0.06 t$ for (b) and $0.08 t$ for (c) around the zero energy, for a linear triple component semimetal with $n=1$. Same set of states, but living within the energy window $\Delta E=0.02 t$ is shown Fig.~\ref{Fig:FermiArc_2D}(a). Note that the seemingly disconnected pieces of the Fermi arc surface states, localized around $k_y=0, \pm \pi$, start to get connected as we increase $\Delta E$. Therefore, these segments belong to the same and one branch of the Fermi arc. This Fermi arc also gets connected with the one on the bottom surface through the bulk triple component points, located at $k_z=\pm \frac{\pi}{2}$, see Fig.~\ref{Fig:FermiArc_3D}(a). The other Fermi arc directly connects two triple component points. Therefore, $n=1$ triple component semimetal accommodates two copies of the Fermi arcs, equal to the monopole charge of the triple component points. Similar outcomes also hold for quadratic ($n=2$) and cubic ($n=3$) triple component fermions. Respectively, three and five Fermi arcs directly connect the corresponding triple component points. The remaining one Fermi arc fragments into two pieces for small $\Delta E$, but ultimately they get connected with increasing $\Delta E$.}~\label{Fig:FermiArc_connected}
\end{figure*}


We begin the discussion with the linear triple-component semimetal. For $k_z=0$, the effective low-energy model of this system is given by 
\begin{equation}\label{Eq:Ham_Chern}
H = k_x S_x + k_y S_y+ \left[ 1-\frac{\alpha}{2} \left( k_x^2 + k_y^2 \right) \right] S_z.
\end{equation}
Even though in the specific tight binding model $\alpha=m=1$ [see Eq.~(\ref{Eq:LatticeModel_Components})], we treat $\alpha$ as a free-parameter in this Appendix. Note that higher-gradient terms proportional to $\alpha$ are irrelevant in comparison to the dominant ${\bf k}$-linear terms at small momentum. However, as we show here such higher gradient terms play paramount important role in properly capturing the topological invariant or the Chern number of the system. At least for $\alpha=1$, the Berry curvature of three bands can be computed analytically, yielding 
\begin{equation}~\label{Eq:Berry_qua}
\Omega^{\pm}_{\mathbf{k}}=\mp\frac{4(2+k_x^2+k_y^2)}{2\pi[4+(k_x^2+k_y^2)^2]^\frac{3}{2}}, \quad \Omega^{0}_{\mathbf{k}}=0,
\end{equation}
from which one can find the Chern number of each band  
\begin{equation}\label{Eq:Chern}
C^{\tau} =  \int^\prime \frac{d^2{\bf k}}{(2 \pi)^2} \: \Omega^{\tau}_{\bf k}.
\end{equation}
In the continuum model the momentum integral is restricted up to an ultraviolet cut-off $\Lambda$ (denoted by the prime symbol in the integral), and we obtain $C^{\pm}=\mp 2$ and $C^0=0$. These numbers do not depend on $\Lambda$. We also numerically compute the Chern number for several other values of the parameter $\alpha$, and find that this number does not depend on $\alpha$, see Fig.~\ref{alpha_c}. Note that in the low-energy Hamiltonian from Eq.~(\ref{Eq:Ham_Chern}), the ${\bf k}$-linear terms dominate at small momentum, whereas the higher gradient terms are more important for large momentum. The term proportional to $S_z$ then plays the role of a ``band-inverted" Wilson mass that ensures the topological nature of the insulating phase. In brief, in the entire construction of topological phases of matter the Wilson mass plays a crucial role, which we further investigate below for triple-component semimetals with $n=2$ and $3$.

The effective low-energy models for quadratic and cubic triple-component semimetals for $k_z=0$ (representing AHIs) can compactly be written as 
\begin{equation}~\label{Eq:TopoHamil_lowenergy}
H_n= d^n_1 ({\bf k}) S_x +d^n_2 S_y ({\bf k}) + \left[ 1- \frac{\alpha}{2} \; f\left( {\bf k}_\perp \right) \right] S_z,
\end{equation}
respectively with $n=2$ and $3$. The functional form of $f\left( {\bf k}_\perp \right)$ depends on the choice of the Wilson mass and the order up to which we expand it in momentum. For example, if we expand $N^2_z({\bf k})$ [see Eq.~(\ref{Eq:LatticeModel_Components})] to quadratic order then $f\left( {\bf k}_\perp \right) = k^2_\perp$ [see Eq.~(\ref{Eq:Ham_Chern})]. The Chern number of one of the topological bands then becomes a function of the parameter $\alpha$, and the correct Chern numbers $\pm 4$ (for $n=2$) are never recovered, see Fig.~\ref{Chern_D_T}(a). Such peculiar outcome roots in the fact that there is no momentum scale separation between $d^2_j({\bf k})$ (for $j=1,2$) and $f\left( {\bf k}_\perp \right)$, and the band-inverted Wilson mass is not capable of capturing the topological invariant of the system.

Let us now consider a different Wilson mass~\cite{roy-goswami-juricic} 
\begin{equation}~\label{Eq:Wilsonmass_New}
N^2_z({\bf k})= m \left[ 6-\sum_{j=x,y} \bigg\{ 4 \cos(k_j)- \cos(2 k_j) \bigg\} \right].
\end{equation} 		
The leading-order expansion of $N^2_z({\bf k})$ around $(k_x,k_y)=(0,0)$ yields $f\left( {\bf k}_\perp \right)= k^4_x+k^4_y$. Irrespective of the coefficient of $f\left( {\bf k}_\perp \right)$, namely $\alpha$, we always find the Chern number of one of the dispersive bands of the underlying two-dimensional AHI to be $+4$, as shown in Fig.~\ref{Chern_D_T}(b).

We find similar outcomes also for the cubic triple-component semimetals. When $f\left( {\bf k}_\perp \right)= k^2_\perp$ the Chern number does not reveal the correct topological invariant of the system, as shown in Fig.~\ref{Chern_D_T}(c). On the other hand, with $f\left( {\bf k}_\perp \right) = k^4_x+k^4_y$ [from the leading order expansion in Eq.~(\ref{Eq:Wilsonmass_New})] we obtain the correct Chern number of the bands, namely $C=+6$ for one of the bands, as shown in Fig.~\ref{Chern_D_T}(d).

Therefore, a continuum model that captures the correct topological invariant must satisfy the following two criteria: (1) The Wilson mass must carry the largest power of momentum (so that it dominates in the ultraviolet regime), and (2) the Wilson mass must change its sign at some momentum (ensuring the band-inversion)~\footnote{We assumed that the phase is topological, not trivial.}. The above discussion along with the results displayed in Fig.~\ref{Chern_D_T} justify the former criterion. In order to appreciate the second one we now present another set of results.

If we expand $N^2_z ({\bf k})$ from Eq.~(\ref{Eq:LatticeModel_Components}) to the fourth and sixth order in momentum we respectively obtain $C=0$ and $+4$ (for $n=2$) or $+6$ (for $n=3$). Expanding $N^2_z ({\bf k})$ from Eq.~(\ref{Eq:LatticeModel_Components}) up to the sixth order in momentum we find 
\begin{equation}
f\left( {\bf k}_\perp \right)=\frac{k^2_x+k^2_y}{2} -\frac{k^4_x+k^4_y}{24} +\frac{k^6_x+k^6_y}{720} + {\mathcal O} \left( k^8_x,k^8_y\right).
\end{equation}
If we keep terms only up to the fourth order in momentum ${\bf k}_\perp$, then it dominates over the quadratic term, but the Wilson mass (proportional to $S_z$ in Eq.~(\ref{Eq:TopoHamil_lowenergy})] does not show band-inversion, and we obtain $C=0$. On the other hand, if we keep terms up to the sixth order in momentum in the above expression, then it dominates in the ultraviolet regime, and also captures the band-inversion. Consequently, we find $C=+4$ (for $n=2$) and $+6$ (for $n=3$). Hence, a low-energy model can only capture all the topological features correctly only when it meets the above mentioned two conditions. On the other hand, in a tight-binding model all higher-gradient terms are present and we always find the correct topological invariant of the system with the Wilson mass $N^2_z ({\bf k})$ introduced in Eq.~(\ref{Eq:LatticeModel_Components}). Irrespective of the choice of the Wilson mass and $f\left( {\bf k}_\perp \right)$ the flat band always possesses exactly zero Chern number.

\section{Connectivity of Fermi arcs}~\label{Appendix:FermiArc_connected}

This Appendix is devoted to establish that the surface localized states near $k_y=\pm \pi$ and $0$, see Figs.~\ref{Fig:FermiArc_2D} and \ref{Fig:FermiArc_3D}, are the segments of \emph{one} Fermi arc. Recall that in Figs.~\ref{Fig:FermiArc_2D} and \ref{Fig:FermiArc_3D}, we show the amplitude of the surface localized states residing within the energy window $\Delta E$ around zero energy. In these two figures they appear as disjoint pieces, due to our choice of the energy window $\Delta E$. However, such seemingly disconnected pieces get connected as we systematically increase the energy window $\Delta E$, as shown in Fig.~\ref{Fig:FermiArc_connected}. For the sake of concreteness, we here show the results only for $n=1$. However, the same conclusion holds for $n=2$ and $3$. To summarize, a triple component semimetal, characterized by triple-component points with monopole charge $2 n$, accommodates $2 n$ copies of the Fermi arc surface states, thus anchoring the bulk-boundary correspondence for spin-1 (or in general any integer spin) topological semimetals.  
\\


\begin{thebibliography}{10}


\bibitem{herring} C. Herring, \emph{Accidental Degeneracy in the Energy Bands of Crystals}, Phys. Rev. {\bf 52}, 365 (1937).

\bibitem{dornhaus} R. Dornhaus, G. Nimtz, and B. Schlicht, \emph{Narrow-Gap Semicounductors}, (Springer-Verlag, Berlin, 1983).

\bibitem{RyuTeo} C.-K. Chiu, J. C. Y. Teo, A. P. Schnyder, and S. Ryu, \emph{Classification of topological quantum matter with symmetries}, Rev. Mod. Phys. {\bf 88}, 035005 (2016).

\bibitem{Barnevig_2016} B. Bradlyn, J. Cano, Z. Wang, M. G. Vergniory, C. Felser, R. J. Cava, B. A. Bernevig, \emph{Beyond Dirac and Weyl fermions: Unconventional quasiparticles in conventional crystals}, Science \textbf{353}, aaf5037 (2016).

\bibitem{kane-prb} B. J. Wieder and C. L. Kane, \emph{Spin-orbit semimetals in the layer groups}, Phys. Rev. B {\bf 94}, 155108 (2016).

\bibitem{hasan-review} M. Z. Hasan, S.-Y. Xu, I. Belopolski, S.-M. Huang, \emph{ Discovery of Weyl Fermion Semimetals and Topological Fermi Arc States}, Ann. Rev. Cond. Mat. Phys. {\bf 8}, 289-309 (2017).

\bibitem{armitage-review}  N. P. Armitage, E. J. Mele, A. Vishwanath, \emph{ Weyl and Dirac Semimetals in Three Dimensional Solids}, Rev. Mod. Phys. {\bf 90}, 15001 (2018).


\bibitem{luttinger} J. M. Luttinger, \emph{Quantum Theory of Cyclotron Resonance in Semiconductors: General Theory}, Phys. Rev. {\bf 102}, 1030 (1956).

\bibitem{kennett-3} B. Roy, M. P. Kennett, K. Yang, V. Juri\v ci\' c, \emph{From Birefringent Electrons to a Marginal or Non-Fermi Liquid of Relativistic Spin-1/2 Fermions: An Emergent Superuniversality}, Phys. Rev. Lett. {\bf 121}, 157602 (2018).

\bibitem{Dora} B. Dor\'{a}, J. Kailasvuori, and R. Moessner, \emph{Lattice generalization of the Dirac equation to general spin and the role of the flat band}, Phys. Rev. B {\bf 84}, 195422 (2011).

\bibitem{Lan-1} Z. Lan, N. Goldman, A. Bermudez, W. Lu, and P. \"{O}hberg, \emph{Dirac-Weyl fermions with arbitrary spin in two-dimensional optical superlattices}, Phys. Rev. B {\bf 84}, 165115 (2011).

\bibitem{liangfu} T. H. Hsieh, J. Liu, and L. Fu, \emph{Topological crystalline insulators and Dirac octets in antiperovskites}, Phys. Rev. B {\bf 90}, 081112 (2014).

\bibitem{manes} J. L. Man\~es, \emph{Existence of bulk chiral fermions and crystal symmetry}, Phys. Rev. B {\bf 85}, 155118 (2012).

\bibitem{cano-bernevig} B. Bradlyn, L. Elcoro, J. Cano, M. G. Vergniory, Z. Wang, C. Felser, M. I. Aroyo and B. A. Bernevig, \emph{Topological quantum chemistry}, Nature (London) {\bf 547}, 298-305 (2017).

\bibitem{neupert-hasan} G. Chang, B. J. Wieder, F. Schindler, D. S. Sanchez, I. Belopolski, S-M. Huang, B. Singh, D. Wu, T-R. Chang, T. Neupert, S-Y. Xu, H. Lin and M. Z. Hasan, \emph{Topological quantum properties of chiral crystals}, Nat. Mater. {\bf 17}, 978 (2018).

\bibitem{grushin_optical} F. Flicker, F. de Juan, B. Bradlyn, T. Morimoto, M. G. Vergniory, A. G. Grushin, \emph{Chiral optical response of multifold fermions}, Phys. Rev. B {\bf 98}, 155145 (2018). 


\bibitem{Chen} C. Chen, S.-S. Wang, L. Liu, Z.-M. Yu, X.-L. Sheng, Z. Chen, and S. A. Yang, \emph{Ternary wurtzite CaAgBi materials family: A playground for essential and accidental, type-I and type-II Dirac fermions}, Phys. Rev. Materials {\bf 1}, 044201 (2017).

\bibitem{Wieder-Kane} B. J. Wieder, Y. Kim, A. M. Rappe, and C. L. Kane, \emph{Double Dirac Semimetals in Three Dimensions}, Phys. Rev. Lett. {\bf 116}, 186402 (2016).


\bibitem{Venderbilt} S. S. Tsirkin, I. Souza, and D. Vanderbilt, \emph{Composite Weyl nodes stabilized by screw symmetry with and without time-reversal invariance}, Phys. Rev. B {\bf 96}, 045102 (2017).

\bibitem{chen-fiete} Q. Chen and G. A. Fiete, \emph{Thermoelectric transport in double-Weyl semimetals}, Phys. Rev. B {\bf 93}, 155125 (2016).

\bibitem{kane-rappe} H. Gao, Y. Kim, J. W. F. Venderbos, C. L. Kane, E. J. Mele, A. M. Rappe, and W. Ren, \emph{Dirac-Weyl Semimetal: Coexistence of Dirac and Weyl Fermions in Polar Hexagonal \textbf{ABC} Crystals}, Phys. Rev. Lett. {\bf 121}, 106404 (2018).

\bibitem{Linagfu-NatPhys} C. Fang, L. Lu, J. Liu and L. Fu, \emph{Topological semimetals with helicoid surface states}, Nat. Phys. {\bf 12}, 936 (2016).




\bibitem{Dai_2016} H. Weng, C. Fang, Z. Fang, and X. Dai, \emph{Topological semimetals with triply degenerate nodal points in $\theta$-phase tantalum nitride}, Phys. Rev. B \textbf{93}, 241202 (2016).

\bibitem{Ding_2017} B. Q. Lv , Z.-L. Feng, Q.-N. Xu, X. Gao , J-Z. Ma , L.-Y. Kong , P. Richard, Y-B. Huang, V. N. Strocov, C. Fang, H-M. Weng, Y-G. Shi, T. Qian, and H. Ding, \emph{Observation of three-component fermions in the topological semimetal molybdenum phosphide}, Nature {\bf 546}, 627 (2017).

\bibitem{Chen_2017} J. B. He,  D. Chen,  W. L. Zhu,  S. Zhang,  L. X. Zhao,  Z. A. Ren, and G. F. Chen, \emph{Magnetotransport properties of the triply degenerate node topological semimetal tungsten carbide}, Phys. Rev. B \textbf{95}, 195165 (2017).

\bibitem{Hasan_2017} G. Chang, S-Y. Xu,  B. J. Wieder,  D. S. Sanchez, S-M. Huang,  I. Belopolski, T-R. Chang, S. Zhang, A. Bansil, H. Lin,  and M. Zahid Hasan, \emph{Unconventional Chiral Fermions and Large Topological Fermi Arcs in RhSi}, Phys. Rev. Lett. \textbf{119}, 206401 (2017).

\bibitem{Zhang_2017} P. Tang,  Q. Zhou, and S-C. Zhang, \emph{Multiple Types of Topological Fermions in Transition Metal Silicides}, Phys. Rev. Lett. \textbf{119}, 206402 (2017).

\bibitem{Soluyanov_2016} Z. Zhu,  G. W. Winkler, Q. S. Wu, J. Li, and A. A. Soluyanov, \emph{Triple Point Topological Metals}, Phys. Rev. X \textbf{6}, 031003 (2016).

\bibitem{Hasan1_2017} G. Chang, S-Y. Xu, S-M. Huang, D. S. Sanchez, C-H. Hsu, G. Bian, Z-M. Yu, I. Belopolski, N. Alidoust, H. Zheng, T-R. Chang, H-T. Jeng,  S. A. Yang, T. Neupert, H. Lin, and M. Zahid Hasan, \emph{Nexus fermions in topological symmorphic crystalline metals}, Sci. Rep. \textbf{7}, 1688 (2017).

\bibitem{Chang_arXiv} C-H. Cheung, R. C. Xiao, M-C. Hsu, H-R. Fuh, Y-C. Lin, and C-R. Chang, \emph{Inducing 3-component fermions in centrosymmetric system by breaking TRS}, arXiv:1709.07763

\bibitem{Chen_2018} J. Li, Q. Xie, S. Ullah, R. Li, H. Ma, D. Li, Y. Li, and X-Q. Chen, \emph{Coexistent three-component and two-component Weyl phonons in TiS, ZrSe, and HfTe}, Phys. Rev. B \textbf{97}, 054305 (2018).

\bibitem{fulga_2017} I. C. Fulga and A. Stern, \emph{Triple point fermions in a minimal symmorphic model}, Phys. Rev. B {\bf 95}, 241116 (2017).


\bibitem{weyl-original} H. Weyl, \emph{Gravitation and the electron}, Proc. Natl. Acad. Sci. U. S. A. {\bf 15}, 323 (1929).

\bibitem{mn3sn} K. Kuroda, T. Tomita, M.-T. Suzuki, C. Bareille, A.A. Nugroho, P. Goswami, M. Ochi, M. Ikhlas, M. Nakayama, S. Akebi, R. Noguchi, R. Ishii, N. Inami, K. Ono, H. Kumigashira, A. Varykhalov, T. Muro, T. Koretsune, R. Arita, S. Shin, Takeshi Kondo, S. Nakatsuji, \emph{Evidence for magnetic Weyl fermions in a correlated metal}, Nat. Mat. {\bf 16}, 1090 (2017).  

\bibitem{ti2mnal} W. Shi, L. Muechler, K. Manna, Y. Zhang, K. Koepernik, R. Car, J. van den Brink, C. Felser, and Y. Sun, \emph{Prediction of a magnetic Weyl semimetal without spin-orbit coupling and strong anomalous Hall effect in the Heusler compensated ferrimagnet Ti$_2$MnAl}, Phys. Rev. B {\bf 97}, 060406 (2018).

\bibitem{lepori} L. Lepori, M. Burrello, E. Guadagnini, \emph{Axial anomaly in multi-Weyl and triple-point semimetals}, JHEP {\bf 06}, 110 (2018).

\bibitem{Gell-Mann} M. Gell-Mann, \emph{Symmetries of Baryons and Mesons}, Phys. Rev. {\bf 125}, 1067 (1962).

\bibitem{nielsen-ninomiya} H. B. Nielsen and M. Ninomiya, \emph{No Go Theorem for Regularizing Chiral Fermions}, Phys. Lett. {\bf 105B}, 219 (1981).

\bibitem{spin-tensor} H. Hu, J. Hou, F. Zhang, and C. Zhang, \emph{Topological Triply Degenerate Points Induced by Spin-Tensor-Momentum Couplings}, Phys. Rev. Lett. {\bf 120}, 240401 (2018).


\bibitem{Fang-HgCrSe} G. Xu, H. Weng, Z. Wang, X. Dai, and Z. Fang, \emph{Chern Semimetal and the Quantized Anomalous Hall Effect in HgCr$_2$Se$_4$}, Phys. Rev. Lett. {\bf 107}, 186806 (2011).

\bibitem{bergevig-MWS} C. Fang, M. J. Gilbert, X. Dai, and B. A. Bernevig, \emph{Multi-Weyl Topological Semimetals Stabilized by Point Group Symmetry}, Phys. Rev. Lett. {\bf 108}, 266802 (2012).

\bibitem{Hasan-PNAS} S-M. Huang, S-Y. Xu, I. Belopolski, C-C. Lee, G. Chang, T-R. Chang, B. Wang, N. Alidoust, G. Bian, M. Neupane, D. Sanchez, H. Zheng, H-T. Jeng, A. Bansil, T. Neupert, H. Lin, and M. Z. Hasan, \emph{New type of Weyl semimetal with quadratic double Weyl fermions}, Proc. Nat. Acad. Sci. {\bf 113}, 1180 (2016). 

\bibitem{nagaosa} B-J. Yang, and N. Nagaosa, \emph{Classification of stable three-dimensional Dirac semimetals with nontrivial topology}, Nat. Commun. {\bf 5}, 4898 (2014).

\bibitem{bera-roy-sau} S. Bera, J. D. Sau, B. Roy, \emph{Dirty Weyl semimetals: Stability, phase transition and quantum criticality}, Phys. Rev. B {\bf 93}, 201302 (2016).

\bibitem{ezawa-2} M. Ezawa, \emph{Chiral anomaly enhancement and photoirradiation effects in multiband touching fermion systems}, Phys. Rev. B {\bf 95}, 205201 (2017).

\bibitem{Xiao:2010} D. Xiao, M. C. Chang, and Q. Niu, \emph{Berry phase effects on electronic properties}, Rev. Mod. Phys. \textbf{82}, 1959 (2010).

\bibitem{calugaru} D. C\u{a}lug\u{a}ru, V. Juri\v ci\' c, and B Roy, \emph{Higher Order Topological Phases: A General Principle of Construction}, Phys. Rev. B {\bf 99}, 041301(R) (2019).

\bibitem{roy-slager-juricic} B. Roy, R-J. Slager and V. Juri\v ci\' c, \emph{Global Phase Diagram of a Dirty Weyl Liquid and Emergent Superuniversality}, Phys. Rev. X {\bf 8}, 031076 (2018).

\bibitem{graphene-RMP} A. H. Castro Neto, F. Guinea, N. M. R. Peres, K. S. Novoselov, and A. K. Geim, \emph{The electronic properties of graphene}, Rev. Mod. Phys. {\bf 81}, 109 (2009).

\bibitem{zhu2018_doubeltriplecomponent} X-Y. Mai, Y-Q. Zhu, Z. Li, D-W. Zhang, and S-L. Zhu, \emph{Topological metal bands with double-triple-point fermions in optical lattices}, Phys. Rev. A {\bf 98}, 053619 (2018).

\bibitem{hasan-kane-review} M. Z. Hasan and C. L. Kane, \emph{Colloquium: Topological insulators}, Rev. Mod. Phys. {\bf 82}, 3045 (2010).

\bibitem{qi-zhang-review} X-L. Qi and S-C. Zhang, \emph{Topological insulators and superconductors}, Rev. Mod. Phys. {\bf 83}, 1057 (2011).

\bibitem{slager} R.-J. Slager, V. Juri\v ci\' c, and B. Roy, \emph{Dissolution of topological Fermi arcs in a dirty Weyl semimetal}, Phys. Rev. B {\bf 96}, 201401 (2017). 

\bibitem{Balents_2017} A. A. Burkov and L. Balents, \emph{Weyl Semimetal in a Topological Insulator Multilayer}, Phys. Rev. Lett. \textbf{107}, 127205 (2011).

 
\bibitem{Sadofyev-1} A. Avkhadiev, and A. V. Sadofyev, \emph{Chiral vortical effect for bosons}, Phys. Rev. D {\bf 96}, 045015 (2017).  

\bibitem{Sadofyev-2}  X-G. Huang, and A. V. Sadofyev, \emph{Chiral Vortical Effect For An Arbitrary Spin}, arXiv:1805.08779

\bibitem{Ziman} J. M. Ziman, \textit{Electrons and Phonons: The Theory of Transport Phenomena in Solids} (Oxford University Press, UK, 2001).

\bibitem{Niu_1999} G. Sundaram, and Q. Niu, \emph{Wave-packet dynamics in slowly perturbed crystals: Gradient corrections and Berry-phase effects}, Phys. Rev. B \textbf{59}, 14915 (1999).

\bibitem{Niu_2006} D. Xiao, Y.Yao, Z. Fang, and Q. Niu, \emph{Berry-Phase Effect in Anomalous Thermoelectric Transport}, Phys.Rev. Lett. \textbf{97}, 026603 (2006).

\bibitem{Son_2012} D. T. Son and N. Yamamoto, \emph{Berry Curvature, Triangle Anomalies, and the Chiral Magnetic Effect in Fermi Liquids}, Phys. Rev. Lett. \textbf{109}, 181602 (2012).

\bibitem{Duval:2006} C. Duval, Z. Horvth, P. A. Horvthy, L. Martina, and P. C. Stichel, \emph{Berry phase correction to electron density in solids and "exotic" dynamics}, Mod. Phys. Lett. B, \textbf{20}, 373 (2006).

\bibitem{Son:2013} D. T. Son and B. Z. Spivak, \emph{Chiral anomaly and classical negative magnetoresistance of Weyl metals}, Phys. Rev. B \textbf{88}, 104412 (2013).

\bibitem{Kim:2014} K-S. Kim, H-J. Kim, and M. Sasaki, \emph{Boltzmann equation approach to anomalous transport in a Weyl metal}, Phys. Rev. B \textbf{89}, 195137, (2014).

\bibitem{Lundgren:2014} R. Lundgren, P. Laurell, and G. A. Fiete, \emph{Thermoelectric properties of Weyl and Dirac semimetals}, Phys. Rev. B \textbf{90} 165115 (2014).

\bibitem{Sharma:2016} G. Sharma, P. Goswami, and S. Tewari, \emph{Nernst and magnetothermal conductivity in a lattice model of Weyl fermions}, Phys. Rev. B \textbf{93}, 035116 (2016).

\bibitem{roy-surowka} R. M. A. Dantas, F. Pe\~{n}a-Benitez, B. Roy, and P. Sur\'owka, \emph{Magnetotransport in multi-Weyl semimetals: A kinetic theory approach}, JHEP {\bf 12}, 069 (2018).

\bibitem{zyuzin} V. A. Zyuzin, \emph{Magnetotransport of Weyl semimetals due to the chiral anomaly}, Phys. Rev. B {\bf 95}, 245128 (2017).





\bibitem{Burkov_2017} A. A. Burkov, \emph{Giant planar Hall effect in topological metals}, Phys. Rev. B {\bf 96}, 041110 (2017).

\bibitem{Nandy_2017} S. Nandy, G. Sharma, A. Taraphder, and S. Tewari, \emph{Chiral Anomaly as the Origin of the Planar Hall Effect in Weyl Semimetals}, Phys. Rev. Lett. \textbf{119}, 176804 (2017).

\bibitem{Nandy1_2017} S. Nandy, A. Taraphder, and S. Tewari, \emph{Berry phase theory of planar Hall effect in Topological Insulators}, Sci. Rep. {\bf 8}, 14983 (2018).

\bibitem{Nandy2_2017}  S. Nandy, A. Taraphder, and S. Tewari, \emph{Planar Thermal Hall Effect in Weyl Semimetals}, arXiv:1711.03102


\bibitem{goswami-roy-dassarma} P. Goswami, B. Roy, and S. Das Sarma, \emph{Competing orders and topology in the global phase diagram of pyrochlore iridates}, Phys. Rev. B {\bf 95}, 085120 (2017). 

\bibitem{andras-roy} A. L. Szab$\acute{\mbox{o}}$, R. Moessner, and B. Roy, \emph{Interacting spin-3/2 fermions in a Luttinger (semi)metal: competing phases and their selection in the global phase diagram}, arXiv:1811.12415

\bibitem{QBT_iridates1} T. Kondo, M. Nakayama, R. Chen, J. J. Ishikawa, E.-G. Moon, T. Yamamoto, Y. Ota, W. Malaeb, H. Kanai, Y. Nakashima, Y. Ishida, R. Yoshida, H. Yamamoto, M. Matsunami, S. Kimura, N. Inami, K. Ono, H. Kumigashira, S. Nakatsuji, L. Balents, and S. Shin, \emph{Quadratic Fermi node in a 3D strongly correlated semimetal}, Nat. Commun. {\bf 6}, 10042 (2015).

\bibitem{QBT_iridates2} M. Nakayama, T. Kondo, Z. Tian, J. J. Ishikawa, M. Halim, C. Bareille, W. Malaeb, K. Kuroda, T. Tomita, S. Ideta, K. Tanaka, M. Matsunami, S. Kimura, N. Inami, K. Ono, H. Kumigashira, L. Balents, S. Nakatsuji, and S. Shin, \emph{Slater to Mott Crossover in the Metal to Insulator Transition of Nd$_2$Ir$_2$O$_7$},  Phys. Rev. Lett. {\bf 117}, 056403 (2016).

\bibitem{Pr2Ir2O7_AHE} Y. Machida, S. Nakatsuji, S. Onoda, T. Tayama, and T. Sakakibara, \emph{Time-reversal symmetry breaking and sponta-
neous Hall effect without magnetic dipole order}, Nature (London) {\bf 463}, 210 (2010).

\bibitem{argyres-adams} P. N. Argyres, and E. N. Adams, \emph{Longitudinal Magnetoresistance in the Quantum Limit}, Phys. Rev. {\bf 104}, 900 (1956).

\bibitem{li-roy-dassarma} X. Li, B. Roy, and S. Das Sarma, \emph{Weyl fermions with arbitrary monopoles in magnetic fields: Landau levels, longitudinal magnetotransport, and density-wave ordering}, Phys. Rev. B {\bf 94}, 195144 (2016).

\bibitem{rahul-spin-1-SC} Y-P. Lin,  and R. M. Nandkishore, \emph{Exotic superconductivity with enhanced energy scales in materials with three band crossings}, Phys. Rev. B {\bf 97}, 134521 (2018).

\bibitem{roy-goswami-juricic} B. Roy, P. Goswami, and V. Juri\v ci\' c, \emph{Interacting Weyl fermions: Phases, phase transitions, and global phase diagram}, Phys. Rev. B {\bf 95}, 201102 (2017). 


\end{thebibliography}
\end{document}